\documentclass[12pt, draftclsnofoot, onecolumn]{IEEEtran}
\usepackage{enumerate}
\usepackage{algorithm}
\usepackage{setspace}
\usepackage[noend]{algpseudocode}
\usepackage{graphics} 
\usepackage{bm}
\usepackage{amsmath}
\usepackage{amssymb}
\usepackage{epstopdf}
\usepackage{nicefrac}
\usepackage{comment}
\usepackage{graphicx}
\usepackage{tabularx}
\usepackage{blkarray}
\usepackage{caption}
\usepackage{subcaption}
\usepackage{amsfonts}
\usepackage{color}
\usepackage{cite}
\usepackage{tikz}
\usepackage[nolist]{acronym}
\graphicspath{{figures/}}
\newtheorem{Remark}{Remark}

\newcommand{\JD}[1]{\textcolor{blue}{#1}}

\begin{acronym}[SPACEEEEEE]
    \acro{5G NR}{5G - New Radio}
	\acro{AWGN}{additive white Gaussian noise}
	\acro{BER}{bit error rate}
	\acro{BPSK}{binary phase shift keying}
	\acro{BP}{belief propagation}
	\acro{CF}{cycle-free}
	\acro{CPM}{circulant permutation matrix}
	\acrodefplural{CPM}[CPMs]{circulant permutation matrices}
	\acro{CN}{check node}
	\acro{CSA}{coded slotted ALOHA}
	\acro{DFT}{discrete Fourier transform}
	\acro{eMBB}{enhanced mobile broadband}
	\acro{ESD}{empirical spectral distribution}
	\acro{FBL}{finite block-length}
	\acro{FEC}{forward error correction}
	\acro{GQ}{generalized quadrangle}
	\acro{i.i.d.}{identical and independent distributed}
	\acro{IoT}{Internet-of-Things}
	\acro{IRSA}{irregular repetition slotted ALOHA}
	\acro{LCR}{low-code rate}
	\acro{LDCD}{Low-density code-domain}
	\acro{LDPC}{low-density parity check}
	\acro{LDS}{low-density spreading}
	\acro{LLR}{log-likelihood ratio}
	\acro{LSD}{limiting spectral distribution}
	\acro{LSL}{large system limit}
	\acro{LSS}{large scale system}
	\acro{LTE}{Long-Term Evolution}
	\acro{MAP}{maximum a-posteriori}
	\acro{MPA}{message passing algorithm}
	\acro{MPF}{marginalize product of functions}
	\acro{MRC}{maximum ratio combining}
	\acro{MAC}{multiple-access channel}
	\acro{ML}{maximum likelihood}
	\acro{MTC}{machine-type communications}
	\acro{mMTC}{massive machine type communications}
	\acro{MMSE}{minimum mean square error}
	\acro{MnAC}{many-access channel}
	\acro{MUD}{multi-user detector}
	\acro{NB}{non-binary}
	\acro{NOMA}{non-orthogonal multiple access}
	\acro{OFDM}{orthogonal frequency division multiplex}
	\acro{QC}{quasi-cyclic}
	\acro{PEG}{progressive edge-growth}
	\acro{PRB}{physical resource block}
	\acro{pmf}{probability mass function}
	\acro{RB}{resource block}
	\acro{RE}{resource element}
	\acro{RS}{random-spreading}
	\acro{CRA}{coded random access}
	\acro{S-EXIT}{scattered extrinsic information transfer}
	\acro{SPARC}{sparse regression code}
	\acro{SNR}{signal-to-noise ratio}
	\acro{SCMA}{sparse coded multiple access}
	\acro{SIC}{successive interference cancellation}
	\acro{SUMF}{single user matched filter}
	\acro{SOTA}{state-of-the-art}
	\acro{URLLC}{ultra-reliable low-latency communication}
	\acro{UE}{user equipment}
	\acro{U-RA}{unsourced random access}
	\acro{VN}{variable node}
	\acro{gNB}{gNodeB}
\end{acronym}

\begin{document}
\title{Sparse Signatures with Forward Error Correction Coding for Non-Orthogonal Massive Access}
\author{Johannes Dommel, 
        Zoran Utkovski, 
        Petar Popovski 
        and S\l awomir Sta\'{n}czak
\thanks{J. Dommel (corresponding author, email: johannes.dommel@hhi.fraunhofer.de), Z. Utkovski and S. Sta\'{n}czak are with the Department of Wireless Communications and Networks, 
Fraunhofer Heinrich-Hertz-Institute, Berlin, Germany. 
S. Sta\'{n}czak is with the Department of Telecommunication Systems, 
Technical University of Berlin, Berlin, Germany. 
P. Popovski is with the Department of Electronic Systems, 
Aalborg University, Aalborg, Denmark.}
}
\maketitle
\begin{abstract}
In massive connectivity scenarios with short packets, of interest is the regime where users share wireless resources in a non-orthogonal fashion. 
Small payloads combined with sporadic user activation call for approaches that jointly address the users’ access to the shared resources and the design of the channel code. 
In this paper, we propose a transmission scheme that combines sparse signatures with finite-length \ac{FEC} coding for non-orthogonal massive access. 
Our signature design is based on Euler squares, which are special instances of quasi-cyclic partial geometries that yield sparse graphs with favorable decoding properties. 
Following a graph-theoretic approach, we explicate the benefits of the coding scheme for the receiver processing that involves joint user detection and decoding. 
The proposed construction is flexible and can be explicitly characterized for a large number of combinations of system parameters, suitable for both grant-based and grant-free massive access. 
Finally, unlike common existing schemes, our scheme can be applied to \ac{U-RA}. 
We numerically characterize the trade-off between system parameters such as number of users, load and channel coding rate. 
The performance evaluation against the state of the art illustrates the potential of the scheme to provide an energy-efficient solution for \ac{U-RA}.
\end{abstract}
%
\begin{IEEEkeywords}
Internet of things (IoT), message passing algorithm (MPA), non-orthogonal multiple access (NOMA), sparse code multiple access (SCMA), unsourced random access (U-RA).
\end{IEEEkeywords}
\acresetall 
\section{Introduction}
\label{sec:introduction}
\IEEEPARstart{T}{he} growing interest in \ac{IoT} applications has put \ac{mMTC} at the focus of the wireless communications research for beyond 5G networks~\cite{Mahmood2020, David18, chen2020massive}.
\ac{mMTC} services are characterized by the presence of a potentially massive number of users that transmit short packets in a sporadic fashion and potential applications are in various domains, ranging from industry and smart cities to logistics and healthcare~\cite{Mumtaz2017, Mehmood2017, Nanda2019, Habibzadeh2020}. 
In these scenarios, of interest is the operational regime where the users share some wireless resources in a \textit{non-orthogonal} fashion~\cite{Ding2017, Chen18, Shin2017, Wan2018, Vaezi2019}. 
Indeed, in massive connectivity scenarios with short messages the traditional orthogonal schemes are expected to be highly  inefficient, mainly due to an inevitable signaling overhead. The problem is exacerbated by the reduced efficiency of channel codes in the \ac{FBL} regime~\cite{polyanskiy2010channel, durisi16, Liva2019Survey}.
Therefore, the design of suitable non-orthogonal access strategies plays a central role in facilitating efficient and effective utilization of scarce wireless resources.
There are two general paradigms for access protocols for massive \ac{IoT}:  grant-based (scheduled) and grant-free (random) radio access. %
In both cases the actual access can be non-orthogonal, over shared resources, for which two central questions arise:
%
%
\begin{itemize}
    \item[\emph{Q1}:] How are the information-carrying messages of individual users assigned to the shared resources? 
    \item[\emph{Q2}:] What is the resulting impact on receiver-side processing, including decoding performance and receiver complexity?
\end{itemize}

%
%
%
%
%

Both are particularly difficult with \emph{grant-free} random access, as the allocation to the shared resources should match the activation statistics of the users, while the decoding process is complicated by the fact that the amount (or even the sheer number) of transmission nodes is unknown.
%
%
While \emph{Q1} was recently answered from an information-theoretical point of view for the \ac{LSS} in favor of a regular sparse mapping~\cite{shental2017low}, practical coding schemes at reasonable complexity are still subject to research, especially for massive connectivity.  
In the context of \ac{NOMA}, the concept of sparsity was introduced in \ac{LDS}~\cite{Hoshyar2008, vdBeek2009, Hoshyar10}, in which modulated constellation points are \textit{repeated} over non-zero elements of a sparse signature. 
It turns out that the sparse structure of the signaling supports a near-optimal receiver implementation based on belief propagation\cite{vdBeek2009} with significantly reduced complexity. 
The concept of \ac{LDS} was generalized within the framework of \ac{SCMA}\cite{Nikopur2013}, which brings an additional shaping gain by mapping information bits directly onto \emph{sparse} codewords.
Further increase in performance at moderate complexity are achieved by exchanging external information between a \ac{MUD} based on \ac{MPA} and a bank of (user-specific) \ac{FEC} decoder, which uses the turbo principle~\cite{Wang1999, Xiao2015,Wu2015, Meng2018} at the receiver.
Considering a number of devices in massive \ac{IoT} in the order of $10^4$ and more, the crucial part of \ac{SCMA} becomes the \emph{codebook} design, which can be broken down into a sparse mapping matrix and a set of complex constellations\cite{taherzadeh2014scma}.  
Therefore, the design problem of \ac{SCMA} codebooks is to jointly optimize sets of complex constellations and mapping matrices according to a given design criterion.
Due to the complexity, sub-optimal multi-stage approaches are considered in which first a sparse mapping and then a set of complex constellations is found, see e.g.~\cite{Vameghestahbanati2019} for a survey on different codebooks designs in the context of \ac{SCMA} systems and their design criteria. 
It reveals, that the performance of a specific codebook design highly depends on the channel type~\cite{Vameghestahbanati2019}, such that a general definition of the design criterion remains elusive.
%
%
%
%
%
%
%
%
%
A joint optimization approach is proposed in~\cite{Peng2017}. The authors propose a heuristic algorithm that aims to minimize the overlap of collided users in every resource block and ensure the largest difference between any two distinct codewords.  
However, due to the nature of the codebook design, \ac{SCMA} typically considers a relatively small number of users sharing a sub-block of the available system resources. 
%
%
In the context of massive access, this requires a certain level of coordination as the system users should be divided in groups, which are then configured on corresponding sub-blocks.

\subsection{Contribution}
\label{sec:contribution}
This paper contributes to the  answer of $\emph{Q1}$ and $\emph{Q2}$ for massive non-orthogonal access as it proposes a transmission scheme that combines the principle of sparse spreading codes 
in combination with user-specific \ac{FEC}. 
The proposed scheme supports a very large number of devices and can be employed for both grant-based and grant-free access. 
Notably, and unlike SCMA or other similar approaches, our scheme can also be applied in the context of  \ac{U-RA}~\cite{polyanskiy17}, a somewhat different radio access paradigm where all users make use of a \textit{shared} (instead of a user-specific) codebook and the receiver decodes the list of transmitted messages irrespective of the identity of the active users.
The main contributions can be summarized as follows: 
\begin{itemize}
    \item We propose a coding scheme for \textit{massive} access based on sparse spreading signatures in combination with channel coding that maps the output of a user-specific \ac{FEC} encoder symbol-wise over a shared set of resources. We propose sparse signature designs based on Euler-squares, which are  special instances of quasi-cyclic partial geometries. Following a graph-theoretic approach, we discuss the implications of the sparse signature design on the joint user detection and decoding process. 
    Unlike \ac{SCMA}, the proposed construction is scalable to a large number of users and shared resources.     
    \item We describe a nearly-optimal, low-complexity  receiver architecture 
    and characterize the performance of the iterative decoding process. 
    For the grant-free access scenario with random user activation, we devise a peeling decoder that further reduces the overall complexity of the multi-user detection and \ac{FEC} decoding steps in the regime of higher user activation.
    \item We numerically evaluate the spectral efficiency of our proposed sparse construction against asymptotic results in the \ac{LSL} and numerically evaluate the joint effects of sparse signature design and \ac{FEC} to characterize the trade-off between system parameters such as sparsity, system load and channel coding rate. 
    %
    \item Finally, we show how the proposed coding scheme can be applied to \ac{U-RA}. The performance evaluation against the state of the art illustrates the potential of the proposed scheme to provide an energy-efficient solution for the unsourced random access scenario with higher user activation. We would like to emphasize that the existing schemes, such as \ac{SCMA}, are not readily extendable to the U-RA setup.
\end{itemize}
%
%
The remainder of this paper is organized as follows.
Section~\ref{sec:context_and_related_work} provides the general context for non-orthogonal massive access and gives a systematic overview on relevant literature.
In Section~\ref{sec:system_model}, we introduce the system model and provide a description of the proposed coding scheme, followed by a description of the iterative receiver architecture. 
In Section~\ref{sec:code_design} we provide the details of the underlying sparse signature design and elaborate on the connection to combinatorial designs based on  partial geometries. We exploit this connection to discuss the impact of the sparse signature design on the overall decoding performance.
In Section~\ref{sec:Results}, we characterize numerically the trade-off between the system parameters such as number of system users, system load, and channel coding rate. 
We also  benchmark the proposed scheme against the state-of-the art schemes in the context of U-RA. 
Finally, Section~\ref{sec:Summary} concludes the paper.
%
%
\section{Context and Related Work}
\label{sec:context_and_related_work}
\subsection{The Context for Non-Orthogonal Massive Access}
While non-orthogonal access has recently attracted considerable attention by both researchers and practitioners~\cite{Ding2017, Chen18, Shin2017, Wan2018, Vaezi2019}, a unifying  framework that structures and classifies the plethora of different approaches investigated in this context is missing. The issue is complex due to the different modelling assumptions that come with different approaches. 
Perhaps the only possible unifying point is in the definition that non-orthogonal access assumes that the same $N$ communication resources are simultaneously used by ($K$) multiple users.  
We describe different models that have emerged from this definition loosely following a historical order in the following. 

\smallskip
\noindent
In ALOHA models~\cite{Abramson1970} for \emph{random access}, a packet is the smallest, atomic unit of information. The focus in the model lies on the fact that the users sending packets are uncoordinated. In the ALOHA model any collision is destructive, which means that non-orthogonality is undesirable. However, non-orthogonality is necessary in order to allow efficient access of uncoordinated users. The ultimate lack of coordination is analyzed by a clever observation that when the number of users $K \rightarrow \infty$, each packet comes from a new user with probability $1$, such that the previous transmissions/collisions of the same users cannot be used to make better scheduling decisions in the future. 

At the other extreme is the information-theoretic treatment of a \ac{MAC} model~\cite{Ahlswede1973}. 
In this model, a group of users is perfectly coordinated in the way they access the channel, so each user is aware at which rate the other users are transmitting.
While the users are coordinated in terms of the protocol, the data sent by each user is neither correlated with the data of the other users nor it is known by the other users. 
The atomic unit in this model is a channel use, in which the base station is a common receiver that gets a noisy version of the combination of codewords sent by all users. 
The main question is what is the capacity region of possible rates that can be selected by the users, so that the receiver (base station) decodes perfectly all users when the codewords become asymptotically long. 
Here it is shown that non-orthogonal use of the \ac{MAC} channel increases the capacity region; this observation has been an inspiration behind a number of works on \ac{NOMA}, e.g. \cite{Rimoldi1996, Zhu2017}.  

From an information-theoretic perspective, non-orthogonal access is related to the conventional \ac{MAC} model that has provided the theoretical background for the study of uplink transmission strategies. Early information-theoretic studies on the \ac{MAC} (see e.g.~\cite{Ahlswede1973}), treat the \ac{MAC} in the so-called ergodic regime where the fundamental limits are studied in the asymptotic limit of infinite coding block-length. In particular, the \ac{MAC} capacity region is computed assuming that the set of transmitting users is typically small  and known in advance and does not change during many channel uses.

Neither of these two models is suited for massive access. The classical collision model in ALOHA does not look into the structure of the packet, while the classical information-theoretic \ac{MAC} model does not deal with uncoordinated access and randomly activated transmissions. Extending over the \ac{MAC} model, random user activation has been integrated in information-theoretic models by way of partially active users (“T-out-of-N MAC”)~\cite{Mathys1990, bar-david93}.
A step towards bridging the gap between the two models is the one used in \ac{CRA} with contention resolution via \ac{SIC}\cite{Casini2007}. In this model for random access, a user does not send her packet only once, as in ALOHA, but she transmits several replicas. A collision of two packets, say $A$ and $B$, is buffered at the receiver as $A+B$ rather than discarded. If there is another replica of $B$ decoded at a different time, then the receiver uses this packet to cancel it from the buffered $A+B$ and thus obtain $A$. In a more advanced version of coded random access, instead of sending replicas of the same packet, the sender sends packets that are related through a \ac{FEC} code~\cite{paolini2015bcoded}. 
\begin{figure}[t!]
    \centering
	\includegraphics[width=0.6\columnwidth]{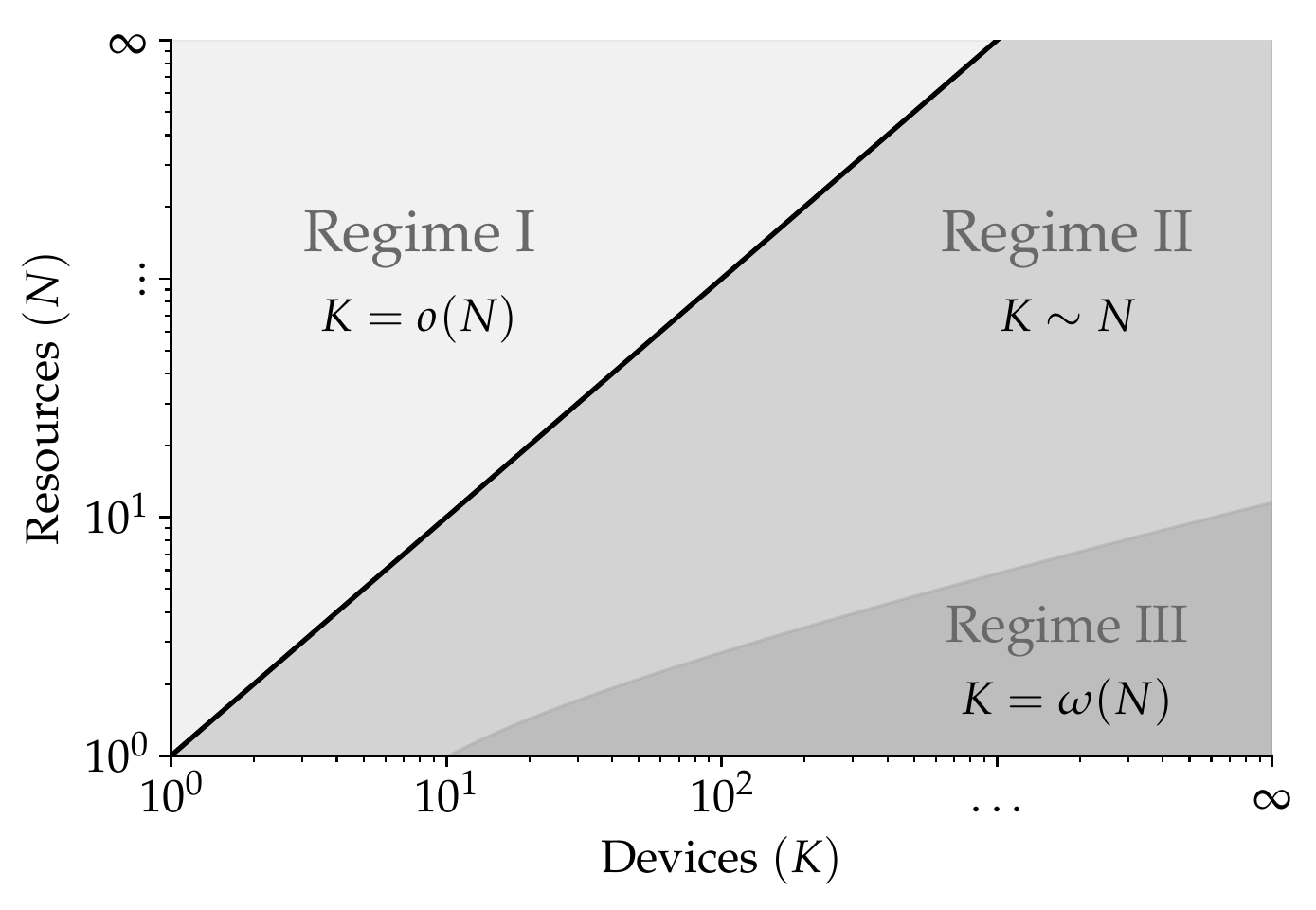}
	\caption{Classification of multiple access models categorized by the asymptotic behaviour of the number of required resources $N$ with the corresponding number of supported devices $K$.}
	\label{fig:OMA_vs_NOMA}
\end{figure}
%

All the models discussed so far are classified within the left upper corner (\emph{Regime I}) of Fig.~\ref{fig:OMA_vs_NOMA}. \ac{CRA} retains the packet as a basic unit, but also looks in the internal structure of the packet and can benefit from a model based on baseband symbols. Yet, the fundamental information-theoretic analysis has still been elusive, as there was a need to find a model that can address both the lack of coordination of users as well as the achievable rates. The setup is complicated by the fact that in massive \ac{IoT} the packets are usually short and the control information, such as the user address, has a size that is comparable to the data size. This was addressed in the information-theoretic model of \ac{MnAC}~\cite{chen14, Chen2017, Ravi2020}, which accommodates random activation as it allows each transmitter to be active with a certain probability in each block. The model solves the problem of encoding control information in an asymptotic regime by allowing the number of users $K$ to increase proportionally to the blocklength $N$. This is in contrast to the standard large-system analysis of multiuser systems in which the blocklength is let to go to infinity before the number of users is made arbitrarily large \cite{shamai97}. We classify the \ac{MnAC} model 
in the middle section (\emph{Regime II}) of Fig.~\ref{fig:OMA_vs_NOMA}.

The problem of transmitting the user address within a finite (short) packet when $K \rightarrow \infty$ was addressed using a completely different approach based on the context of \ac{U-RA}~\cite{polyanskiy17}.  
Under this framework, the users employ the same codebook and collisions are interpreted as the event where multiple users transmit the same codewords. 
In this scenario, the problem of user identification is separated from the actual data transmission, and the decoder only declares which messages were transmitted, without associating the messages to the users that transmitted them. 
The goal of the decoder is to output a list that should contain the different messages that were transmitted by the active users. Under this framework, a collision is the event where multiple users transmit the same codewords from the joint codebook. We note that this formulation requires that the number of active users is known to the decoder, but also to the accessing users, so that they can select the optimal codebook. In addition, the error probability does not account for false  alarms,  i.e.  for  the error  events  when  the  list produced by the decoder contains messages that have not been transmitted by any of the users. As a consequence, the total number of users $K$ can be left out of the model, i.e. it can increase without bound with the blocklength $N$. We classify this model and the respective transmission schemes (see, e.g., \cite{ordentlich2017low}) in the lower-right corner (\emph{Regime III}) of  Fig.~\ref{fig:OMA_vs_NOMA}.

Finally, from a practitioner's point of view, non-orthogonal access assumes the same \ac{PRB} to be shared simultaneously by more than one user, in either scheduled or in a random (grant-free) fashion. 
For example, in an \ac{OFDM} - based system as, e.g., \ac{LTE} and \ac{5G NR}, one \ac{PRB} encompasses $N = n_{sc}\cdot n_o$ \acp{RE}, where $n_{sc}$ is the number of subcarriers and $n_o$ is the number of \ac{OFDM} symbols that form the \ac{PRB}\footnote{Typical values, e.g. in \ac{LTE}, are $n_{sc}=12$ and $n_o=14$.}, and represents the fundamental instance over which users can be scheduled or configured. 
In a more general setting, non-orthogonal access also includes the scenario where $K$ users may be simultaneously scheduled or configured on $N_{PRB} < K$ \acp{PRB}. 
We note that the model assumes that  even in the asymptotic regime the blocklength $N$ is sent to infinity before the number of users $K$ is sent to infinity\cite{shamai97, Ahlswede1973}. Hence, in principle, orthogonal operation is possible, although it might be sub-optimal in general. 
Most of the communication schemes targeting \ac{eMBB} transmissions, investigated under the framework of \ac{NOMA} fall into this category, see \cite{Ding2017} for an overview. 
We classify this model (and the respective transmission schemes) in the upper-left corner (\emph{Regime I}) of Fig.~\ref{fig:OMA_vs_NOMA}.
%

%
Based on this context, we can position our proposed scheme primarily as one that works with baseband symbols and uses them to convey user identification and data.
It can deal with uncoordinated users, while the use of \ac{FEC} allows reaping gains that stem from the information-theoretic models discussed above. In addition, we show how can the proposed approach be adapted to design an unsourced access scheme. 
%
%
\subsection{Related Work}
%
%
%
\subsubsection{Sparse-Coded Multiple Access}
Some of the relevant aspects discussed in this paper (see Section~\ref{sec:contribution}) have been studied in the context of \textit{signature design} for \ac{LDS}, where users are multiplexed on \acp{PRB} by applying signatures as patterns for accessing the shared resources, typically concatenated with low-rate error-correcting codes. 
%
%
The signatures, i.e. the mapping between users and resources, can be either \textit{regular}, where each user occupies a fixed number of resources, and each resource is used by a fixed number of users; or \textit{irregular}, where the respective numbers are random, and only fixed on average. 
In the \ac{LSL}, the optimal spectral efficiency of irregular constructions was shown in \cite{yoshida2006analysis} to reside below the well-known spectral efficiency of dense \ac{RS} \cite{verdu1999spectral}, which stems from the random nature of the user-resource mapping, due to which some users may end up without any designated resources, while some resources may be left unused. 
Recently it was shown that, in certain operational regimes, low-density spreading signatures for \ac{NOMA} based on regular-sparse constructions outperform not only irregular-sparse, but also dense constructions, both for Gaussian channels \cite{shental2017low, Zaidel2018}, and for block-fading channels\cite{Ferrante18}. 
This observation has important practical implications, as sparse mappings allow for feasible near-optimal (\ac{MPA}-based) multiuser receiver implementation where the \ac{BP} benefits from the sparsity of the corresponding factor graph. This is in contrast to dense spreading sequences \cite{verdu1999spectral}, where the complexity of the optimum receiver may be prohibitive for large systems. 
One design proposed for \ac{LDS}~\cite{Hoshyar2008} utilizes a class of sequences derived from regular \ac{LDPC} codes following Gallager's method~\cite{GallagerLDPC}.  
This design is based on regular constructions with random sub-matrices and hence does not guarantee that small cycles are not present (we will discuss these effects in more detail in Section~\ref{sec:code_design}). 
In order to construct sparse signatures and to avoid the occurrence of small cycles, an iterative graph based method based on \ac{PEG}\cite{XIAOHU2005}  
was adopted in~\cite{Qi2017} in order to maximize the local girth at the current variable node, where edges are added to the factor graph progressively in an edge-by-edge manner.
With the objective to maximize the sparsity, a multi-objective programming problem is formulated in~\cite{Qi2017b} that simultaneously maximizes the sum rate and sequences sparsity by using frame theory.
However, optimal sequences for \ac{NOMA} that can achieve both the maximum sum rate and maximum sparsity are still unknown. 
\subsubsection{Coded Random Access}
Aspects of deterministic sparse mappings have also been studied in the literature on \ac{CRA} with the note that in that case \emph{coding} is usually performed at the level of packets~\cite{liva11}, rather than on symbol level and the signature can be interpreted as repetition pattern.  
The performance of random versus deterministic repetition coding is compared in \cite{Boyd2019} where the results show that a deterministic coding approach can lead to a significantly superior performance when the arrival rate is low.
In \cite{Ustinova2019}, a low complexity scheme based on T-fold ALOHA with \ac{SIC} procedure is proposed which bridges the gap between coded random access and \ac{U-RA}. 
According to this approach, a concatenated code construction with outer \ac{NB}-\ac{LDPC} code and inner linear binary code is decoded each slot with an iterative joint decoding algorithm. 
However, besides adequate performance, the main disadvantage is the exponential growth in decoding complexity as the \ac{NB}-\ac{LDPC} code field size increases.
\subsubsection{Unsourced Random Access (U-RA)}
Recent schemes proposed for \ac{U-RA} use sparse mappings, i.e. a structure according to which fractions of a codeword are transmitted in a slotted fashion. 
One coding scheme for the unsourced Gaussian random access channel was proposed in~\cite{ordentlich2017low}. 
According to this approach, the available channel uses are split into sub-blocks of equal length, and each active user randomly chooses only one of these sub-blocks, over which it transmits. 
All users encode their messages using the same codebook obtained by concatenation of two codes: (\emph{i}) an inner binary linear code, whose goal is to enable the receiver to decode the modulo-2 sum of all codewords transmitted within the same sub-block and (\emph{ii}) an outer code, the aim of which is to enable the receiver to recover the individual messages that participated in the modulo-2 sum. 
An alternative state-of-the-art scheme for the \ac{U-RA} channel was proposed in \cite{Fengler2019} based on \acp{SPARC}.
Accordingly, each user encodes its message into a sparse binary vector, which is then mapped to the transmitted signal vector by performing linear mapping using a shared dictionary which is divided into sections. 
An outer code is used to assign the symbols to individual messages. It has been shown in~\cite{Fengler2020}, that this scheme can achieve a vanishing per-user error probability in the limit of large blocklength and a large number of active users at sum-rates up to the symmetric Shannon capacity.
In \cite{Amalladinne2020}, a coded compressed sensing approach is proposed for unsourced multiple-access communication. 
Each active device divides its data into several sub-blocks and then adds redundancy using a systematic linear block code.
%
%
Numerical results demonstrate that coded compressed sensing outperforms other existing practical access strategies over a range of operational scenarios.

\section{System Model and Coding Scheme}
\label{sec:system_model}
\subsection{System Model and Coding Scheme}
Consider a Gaussian \ac{MAC} where a set $\mathcal{K}$ of $K$ users in total attempt to access a block of $N = n_s \ell$ orthogonal resources (or \textit{channel uses}). 
If a user $k$ is active (scheduled or at random), a codeword $\boldsymbol{X}_k \in \mathbb{C}^{ n_s \times \ell}$ is selected for transmission and the received signal reads
\begin{align}
\boldsymbol{Y} 	&= \sum_{k \in  \mathcal{K}} a_k \sqrt{P_k} \boldsymbol{X}_k + \boldsymbol{W}, \label{eq:y_rx}
\end{align}
where $\boldsymbol{a}= [a_1, a_2, \ldots, a_K] \in \left\{0,1\right\}^K$ is the "activity vector"; $P_k$ is the transmit power (possibly also incorporating path-loss effects) of user $k$; the elements of $\boldsymbol{W}\in\mathbb{C}^{n_s \times \ell}$ are i.i.d. $\mathcal{CN}(0, 1)$; and  $\boldsymbol{Y}\in\mathbb{C}^{ n_s \times \ell}$.
We assume the same power level for all users ($P_k = P, \forall k \in \mathcal{K}$), and define the  \textit{per-user}~\ac{SNR} $ \triangleq P$. 
%
%
We note that, in the case of \emph{scheduled access}, the activity pattern is known to the receiver. In the grant-free scenario with random activation, on the other hand, we assume that the receiver knows only the statistics of $\boldsymbol{a}$, but not its' realization. In \ac{U-RA} there is the additional condition, that all the codewords are drawn from the same codebook. 
%
\subsection{Block-sparse Coded Modulation}
\label{sec:coding_scheme}
\begin{figure}[bht]
    \centering
	\includegraphics[width=0.6\columnwidth]{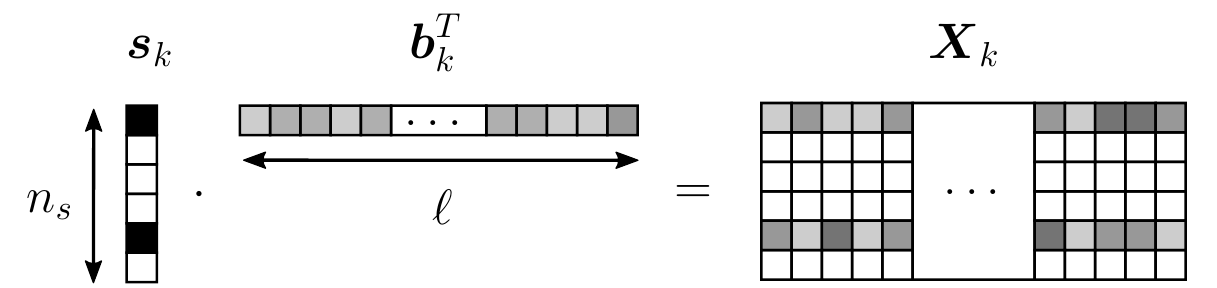}
	\caption{Illustration of the coding scheme: the outer product of a finite-length codeword $\boldsymbol{b}_k$ and a sparse signature $\boldsymbol{s}_k$ constitute the block-sparse codeword $\boldsymbol{X}_k$ of size $N = n_s \cdot \ell$.}
	\label{fig:LDSTxChain}
\end{figure}

The encoder is defined for each user as follows. 
The $N$ channel uses are split into $n_s$ sub-blocks of length $\ell= N/n_s$. 
%
%
When active, i.e. $a_k = 1$, user $k$ selects a message $w_k$ from its message set ${\mathcal{W}_k\triangleq \{1,\ldots, M_k\}}$ and maps it to a binary codeword ${\Delta_k\in \mathbb{F}_2^{n}}$ from a \ac{FEC} code $\mathcal{C}_{\mathrm{FEC}}^{(k)}$. 
While in general it is possible that each user applies a different \ac{FEC} code (which can also be non-linear), in the following we will assume that all users apply the same linear block code $\mathcal{C}_{\mathrm{FEC}}$ of rate $R_{FEC} = \log_2 |\mathcal{W}|/n$, where $\mathcal{W}\triangleq\mathcal{W}_1=\cdots=\mathcal{W}_K$ is the message set of all users. 
The encoded block of bits $\Delta_k$ is interleaved $\Delta_k'  = \pi_k(\Delta_k)$ in order to randomize dependencies between bits and modulated to a complex vector of constellations $\boldsymbol{b}_k$ at modulation order $Q_k = \ell / n$, subject to the power constraint ${\|\boldsymbol{b}_k\|^2_2\leq \ell}$. 
Finally, the vector $\boldsymbol{b}_k$ is mapped onto the $n_s$ sub-blocks to obtain the codeword $\boldsymbol{X}_k=\boldsymbol{s}_k \boldsymbol{b}_k^T \in \mathbb{C}^{n_s \times \ell}$, as illustrated in Fig.~\ref{fig:LDSTxChain}, where $\boldsymbol{s}_k$ is chosen as the $k$-th column of a signature matrix $\boldsymbol{S}\in\mathbb{C}^{n_s \times K}$. 
The  signature matrix 
is defined as $\boldsymbol{S} \triangleq \boldsymbol{F}\boldsymbol{D}^{-1}\boldsymbol{\Phi}$, with $\boldsymbol{F} = [\boldsymbol{f}_1, \ldots, \boldsymbol{f}_K] \in \{0, 1\}^{n_s \times K}$ being a sparse matrix; 
${\boldsymbol{D} = \text{diag}\left(\|\boldsymbol{f}_1\|, \ldots, \|\boldsymbol{f}_K\|\right)}$ with 
$\|\cdot\| \triangleq \sqrt{\langle\cdot, \cdot \rangle}$ and $\boldsymbol{\Phi} = \text{diag}(\phi_1, \ldots, \phi_K)$ with $\{\phi_k\}_{k\in \mathcal{K}} \in \mathbb{C}$ being phase-rotations~\cite{taherzadeh2014scma} on the complex union circle.
%
%
\begin{Remark}
We note that for \ac{U-RA}, the users pick randomly a column index from the shared signature matrix for transmission, such that $\boldsymbol{X}_k=\boldsymbol{s}_i \boldsymbol{b}_k^T$ with $i$ chosen uniform \ac{i.i.d.} on [K].\label{remark:ura}
\end{Remark}

\smallskip
\noindent
The common decoder assigns a decision $\hat{\boldsymbol{w}} = \{\hat{w}_{k}\}_{k \in \mathcal{K}}$ after observing $N$ channel uses~\eqref{eq:y_rx}, where ${\hat{w}_k \in \{0 \cup \mathcal{W}\}}$, with $\hat{w}_k = 0$, if the decoder decides in favor of user $k$ being \emph{inactive}. 
The following error events are defined for the different access setups: in the scheduled case, an error event occurs for user $k$, if the codeword is not decoded correctly, which is defined as $E^k_1 \triangleq \{\hat{w}_k \neq w_k\}$.
In the random access case, additionally to $E^k_1$, an error event occurs if an active device is not detected,  which is defined as $E^k_2 \triangleq \{\hat{w}_k = 0~\vert~a_k = 1\}$. 
For \ac{U-RA}, in addition to $E^k_1$ and $E^k_2$, a third error event $E^k_3$ occurs if two users choose the same signature for transmission, defined as $E^k_3 \triangleq \{\boldsymbol{s}_k = \boldsymbol{s}_j~\text{for any}~j \neq k~\vert a_j = 1\}$.
We note that the code achieves an energy-per-bit to noise power spectral density ratio $E_b / N_0 \triangleq \frac{\ell}{2 \log_2 |\mathcal{W}|}\cdot \text{SNR}$ per complex dimension with effective code rate $R = \frac{\log_2 |\mathcal{W}|}{n_s\cdot \ell} = \frac{R_{FEC}}{n_s}$ bits per channel use.
\subsection{Receiver Processing}
\label{sec:Receiver_Processing}
\begin{figure}[hbt]
\centering
\includegraphics[width=0.6\columnwidth]{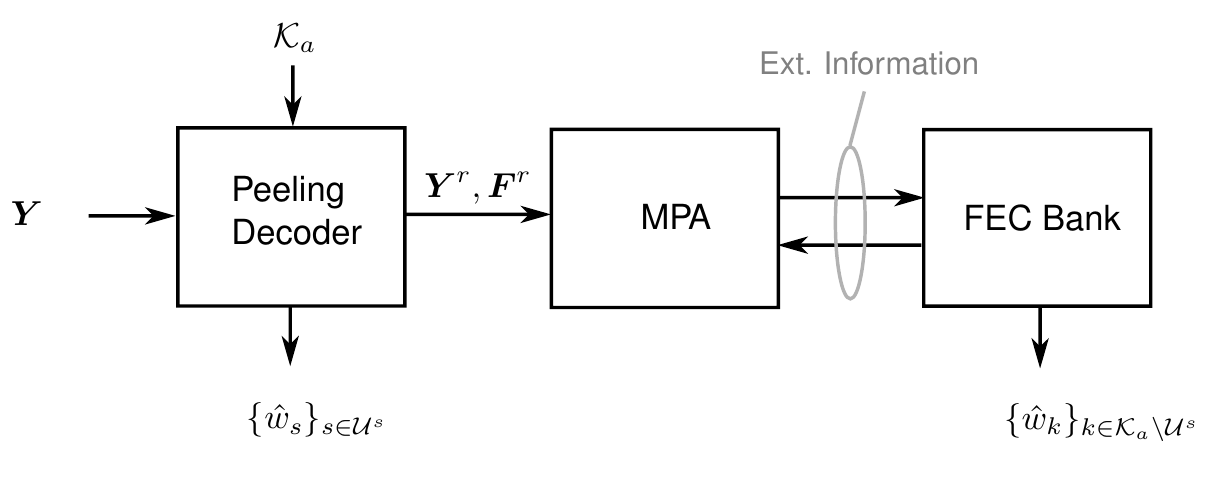}
\caption{Receiver structure employing a peeling decoder concatenated with a turbo-enhanced \acf{MPA}.}\label{fig:PeelingTurboRx}
\end{figure}

Under the assumption that the set of \textit{active} devices ${\mathcal{K}_a  = \{k \in \mathcal{K} \vert a_k=1\}}$ is known  a-priori to the receiver (e.g. in the scheduled case), the optimum multiuser detection problem can be solved by maximizing the joint a-posteriori \ac{pmf} of all transmitted symbols, i.e. to estimate ${\hat{\boldsymbol{w}}_a =  \{\hat{w}_k\}_{k \in \mathcal{K}_a}}$ that maximizes the posterior $p_{\mathbf{w}_a \vert \mathbf{Y}}$ of all transmitted symbols ${\boldsymbol{w}_a =  \{w_k\}_{k \in \mathcal{K}_a}}$ given the received signal~\eqref{eq:y_rx} and the activity pattern $\boldsymbol{a}$. 
Hence, the \ac{MAP} detector can be expressed as 
\begin{align}
    \hat{\boldsymbol{w}}_a &= \arg \max\limits_{\boldsymbol{w}_a \in \mathcal{W}^{K_a}} p_{\mathbf{w}_a \vert \mathbf{Y}} (\boldsymbol{w}_a \vert \boldsymbol{Y}, \boldsymbol{a}),  \label{eq:map}  
\end{align}
which quickly becomes computationally prohibitive as the complexity grows exponentially with $K_a = \lvert\mathcal{K}_a \rvert$ due to the evaluation of $\lvert \mathcal{W}\rvert^{K_a}$ signal alternatives. 
While exact computation of the posterior is generally intractable, a low-complex approximation,  introduced in a previous work~\cite{Dommel19}, is summarized in the following. 
The proposed receiver operates on the \textit{pruned} factor graph using a concatenation of (\emph{i}) a peeling decoder, and (\emph{ii}) an \ac{MPA}-based \ac{MUD} in combination with a bank of \ac{FEC} - decoder, as illustrated in Fig.~\ref{fig:PeelingTurboRx}.

Peeling decoding has been well established in coding for the erasure channel and found applications in the the context of compressed sensing\cite{Luby2001, Liu2019, zeng2016peeling, Li19}.  
In contrast to sparse recovery applications, we deploy the peeling decoder principle to decode and remove block-wise messages from the received signal. 
The procedure is summarized as follows: 
Depending on the active set $\mathcal{K}_a$ (which defines the connectivity of the sparse bipartite graph), the received signals (observations) associated with the check nodes can be categorized into: (\emph{i}) \textit{zero-ton} check nodes which does not involve any non-zero symbols; (\emph{ii}) \textit{single-ton} check nodes which involve only one non-zero symbol and (\emph{iii}) \textit{multi-ton} check nodes whose value is the sum of more than one non-zero symbol. 
We note, that the receiver may differentiate (\emph{i})-(\emph{iii}) based on the knowledge of the full factor graph and the \emph{active set} $\mathcal{K}_a$.
With definition of the corresponding incidence matrix $\boldsymbol{F}(\mathcal{K}_a)$, the peeling decoder principle is to successively decodes messages from devices (variable nodes) which are connected to the single-tons. 
To be specific, let $\mathcal{V}^r$ and $\mathcal{C}^r$ denote a set of variable - and check nodes, respectively, and  
let 
$\mathcal{V}^r_{s} \subset \mathcal{V}^r$ denote the set of variable nodes (users) which are at least connected to one single-ton check-node (which can be directly obtained by evaluating the pruned incidence matrix). 
We note that each check node in the factor-graph corresponds to a length-$\ell$ block of \acp{RE}.
The peeling process decodes and removes the signals from users which are connected to single-ton check nodes as described in Algorithm~\ref{alg:peeling_dec}.
{\small
\begin{algorithm}
\caption{Peeling Decoder}\label{alg:peeling_dec}
    \begin{algorithmic}[1]
        \State \textbf{input}~$\boldsymbol{Y}, \boldsymbol{F}, \boldsymbol{a}$
        \State $\mathcal{K}_a^r \gets \mathcal{K}_a$
        \State $\boldsymbol{Y}^r \gets \boldsymbol{Y}$ 
        \State identify $\mathcal{V}^r_{s}$ from $\mathcal{K}_a^r$
        \State $\mathcal{U}^s \gets \emptyset$ \Comment{list of decoded messages}
        \While{$\mathcal{V}^r_{s} \neq \emptyset$}
            \State decode $\{\hat{w}_k\}_{k \in \mathcal{V}_s^r}$ 
            \State $\boldsymbol{Y}^r \gets \boldsymbol{Y}^r - \sum_{k \in  \mathcal{V}_s^r} \boldsymbol{s}_k \hat{\boldsymbol{b}}_k^T$
            \State $\mathcal{K}_a^r \gets \mathcal{K}_a^r \setminus \mathcal{V}^r_{s}$
            \State update $\mathcal{V}^r_{s}$ from $\mathcal{K}_a^r$
            \State $\mathcal{U}^s \gets \mathcal{U}^s \cup \mathcal{V}_s^r$
        \EndWhile
    \end{algorithmic}
\end{algorithm}
}

After completion, the peeling decoder outputs a list of decoded messages $\mathcal{U}^s$ and forwards the residual factor graph (described by the incidence matrix $\boldsymbol{F}^r$) and the "peeled" signal $\boldsymbol{Y}^r$ to an iterative joint \ac{MUD} which operates on the underlying (residual) factor graph on symbol-level and exchanges extrinsic (soft-) information between a bank of (user-specific) \ac{FEC} exploiting turbo-principle~\cite{Wang1999, Xiao2015, Wu2015, Meng2018} on \ac{FEC}-block level. 
%
%
\subsubsection*{Activity Detection}
When the set of active users $\mathcal{K}_a$ (equivalently the structure of the pruned graph) is not known \emph{a-priori}, it needs to be estimated in some way and this is the realistic case under which the algorithm operates.
%
%
%
With sufficiently sparse signatures, this problem is essentially one of sparse support recovery in compressed sensing for which theoretical analysis has been presented in~\cite{Li19}, where a scheme based on sparse graph codes was proposed. 
The scheme provably achieves an \textit{order-optimal scaling} in both the measurement cost and the computational run-time in the \emph{sub-linear} sparsity regime.   
By considering the overhead (in terms of channel resources) associated with the recovery of the active set, one can (roughly) claim an effective decrease of the spectral efficiency $\eta$ by a factor of $\frac{N-\mathcal{O}(K_a  \log (K/K_a))}{N}$. 
\begin{Remark}
We note that this overhead will be moderate in the scenario of interest where the expected number of active users is sub-linear in the total number of system users, $K_a=\mathcal{O}(K^{\delta})$, for some $0<\delta<1$, but the number of available resources scales linearly with the total number of system users, $N=\mathcal{O}(K)$ with fixed system load $K/N$. An optimal design of the measurement process for the estimation of the active set of users is out of the scope of this paper and is left for future consideration.
\end{Remark}

\subsection{Example}
\begin{figure}[t!]
\centering
\begin{subfigure}[c]{0.36\columnwidth}
\includegraphics[width=\columnwidth]{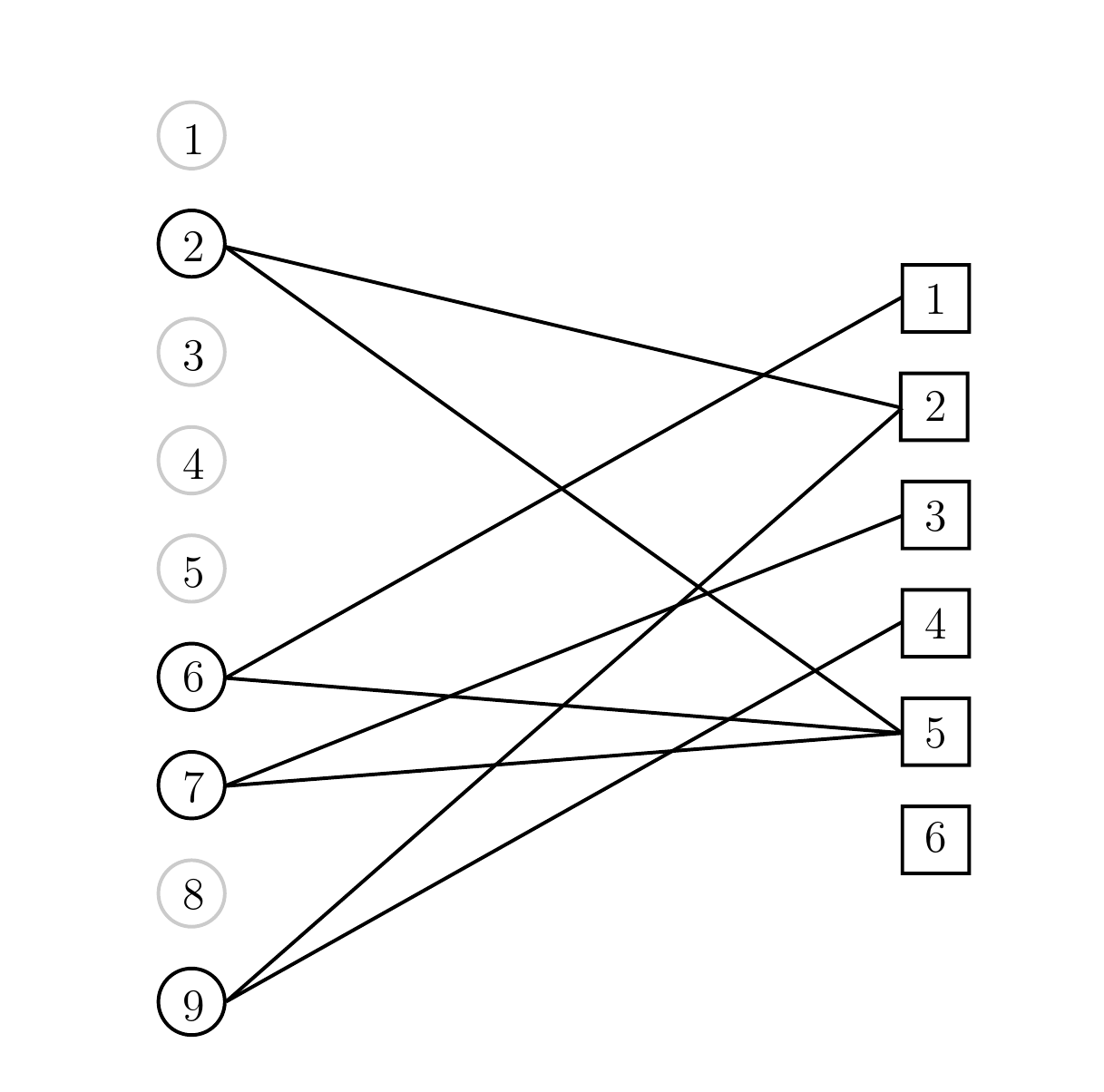}
\caption{Pruned factor graph}\label{fig:FactorGraph_pruned_p0}
\end{subfigure}
\begin{subfigure}[c]{0.36\columnwidth}
\includegraphics[width=\columnwidth]{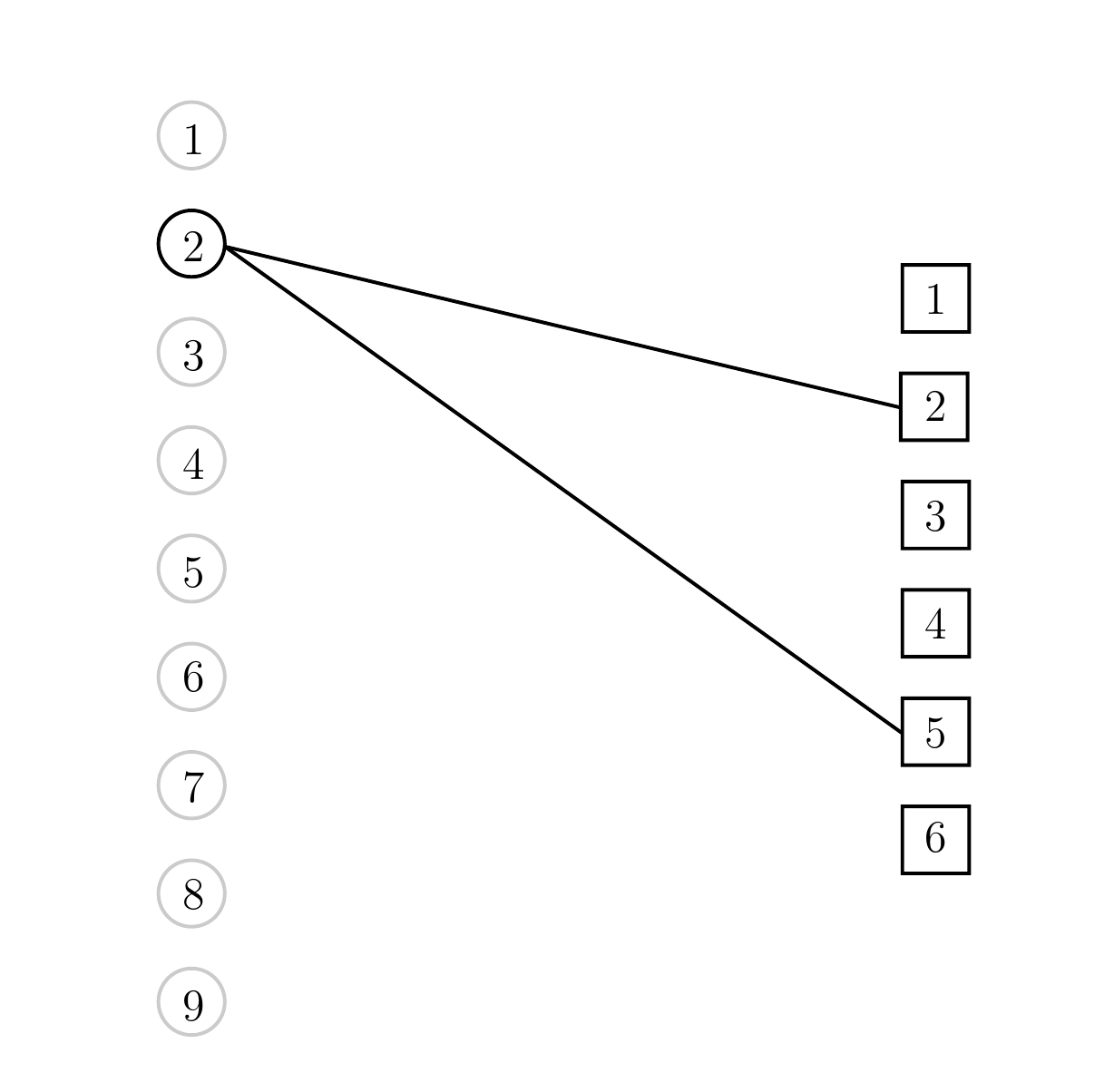}
\caption{After first peeling step}\label{fig:FactorGraph_pruned_p1}
\end{subfigure}
\caption{Peeling decoding example with $\mathcal{K}_a = \{2,6,7,9\}$.}
\label{fig:peeling_example}
\end{figure}

For illustration, consider the following example where $9$ users share $6$ resources, as prescribed by the sparse matrix $\boldsymbol{F}$, defined as 
{\small
\begin{align} 
\boldsymbol{F}&=\left(\begin{array}{ccccccccccc} 
1 & 0 & 0 & | & 0 & 0 & 1 & | & 0 & 1 & 0\\ 
0 & 1 & 0 & | & 1 & 0 & 0 & | & 0 & 0 & 1\\ 
0 & 0 & 1 & | & 0 & 1 & 0 & | & 1 & 0 & 0\\ 
- & -  & -  & | & - & - & -   & | & -  & - & - \\
1 & 0 & 0 & | & 0 & 1 & 0 & | &  0 & 0 & 1\\ 
0 & 1 & 0 & | & 0 & 0 & 1 & | &  1 & 0 & 0\\ 
0 & 0 & 1 & | & 1 & 0 & 0 & | &  0 & 1 & 0 
\end{array} \right) \label{eq:matrix_example}.
\end{align}
}
The construction in (\ref{eq:matrix_example}) is derived from the Euler square $E(3,2)$ (we discuss the specifics of the construction in Section~\ref{sec:code_design}). The mapping is biregular, such that the information from each user is mapped on exactly $2$ resources, and exactly $3$ users overlap on each of the available resources. 
In addition, two rows (columns) of $\boldsymbol{F}$ have at most one place where both have non-zero entries, i.e. there is an overlap on mostly one position. Moreover, we observe that %
$\boldsymbol{F}^{\mathrm{T}}$ is a $3 \times 2$ array of \acp{CPM}, each of size $3 \times 3$. 
We note that the matrix $\boldsymbol{F}$ is the incidence matrix of a bipartite graph that prescribes the message passing procedure at the receiver and each column of $\boldsymbol{F}$ corresponds to the pattern by which the individual users accesses the wireless resources. 
As we will explain in more detail below, the above sparse design characteristics are beneficial to the decoding process performed on the corresponding bipartite graph in both the grant-based and grant-free scenarios.
Consider for example the grant-free scenario with random activation, where only a subset of devices $\mathcal{K}_a = \{2, 6, 7, 9\}$ is active. In that case the receiver operates on the \emph{pruned} graph, where the variable nodes and edges associated with the \textit{inactive} users, i.e. the corresponding columns in $\boldsymbol{F}$, are removed. 
In this case, 
a peeling decoder can be employed to reduce the complexity of the message passing procedure. 
The corresponding (pruned) factor graph is depicted in Fig.~\ref{fig:FactorGraph_pruned_p0}, where factor node $6$ is a zero-ton, the set of factor nodes $\{1, 3, 4\}$ and $\{2, 5\}$ are are single- and multi-tons, respectively.
Hence, the initial set of factor nodes which are connected to single-ton factor nodes can be obtained as $\mathcal{V}^r_{s} = \{6,7,9\}$ by evaluating the structure of the (pruned) incidence matrix.
%
The residual factor graph after the first iteration of the peeling decoder is depicted in Fig.~\ref{fig:FactorGraph_pruned_p1}. 
It can be observed that only two iterations are required to completely decode all active devices.  

\section{Sparse Signature Design}
\label{sec:code_design}
In the following we present the details of the sparse code design based on the concept of Euler-squares~\cite{macneish1922euler} 
which is motivated by recent works on deterministic binary matrices for compressed sensing
\cite{Naidu2016}. 
From combinatorics and experimental design, an Euler square $E(\gamma, \rho)$ of order $\gamma$ and degree $\rho$ consist of $\rho$ mutual orthogonal Latin squares\footnote{The name "Latin square" was inspired by Leonhard Euler (1707 - 1783), who used Latin characters as symbols~\cite{wallis2016introduction}.} of size $\gamma \times \gamma$. A Latin square is a $\gamma \times \gamma$ array filled with $\gamma$ different symbols, each symbol occurring exactly once in each row and in each column, respectively. 
Consider for example an Euler square $E(3,2)$ of order $\gamma = 3$ degree $\rho = 2$, which can be written as:
%
{\small
\begin{equation}
  \begin{array}{ccc} 
(1,1) & (2,2) & (3,3) \\ 
(2,3) & (3,1) & (1,2) \\ 
(3,2) & (1,3) & ~(2,1). 
\end{array}
\end{equation}
}
Based on an Euler square $E(\gamma, \rho)$, a binary sparse matrix $\boldsymbol{F} \in \{0,1\}^{\gamma\rho \times \gamma^2}$ can be constructed with the following properties:
\begin{enumerate}[(I)]
    \item $\boldsymbol{F}$ is biregular, with $\rho$~non-zero entries in each row, and $\gamma$ in each column;\label{es_prop_1}
    \item Two rows (columns) of $\boldsymbol{F}$ have at most one place where both have non-zero entries (i.e. there is an overlap on one position at most. Matrices of this type are called \textit{row and column} (RC) constrained;\label{es_prop_2}
    \item $\boldsymbol{F}^{T}$ is a $\gamma \times \rho$ array of $\gamma \times \gamma$ \acp{CPM}.\label{es_prop_3}
\end{enumerate}
Compared to general biregular mappings, property~(\ref{es_prop_2}) and ~(\ref{es_prop_3}) offer additional advantages for the decoding process both in the case of scheduled and grant-free transmissions. 
For example, from~(\ref{es_prop_2}) it follows that any two users will overlap on one resource at most.  
For analogous discussions on the advantages of general combinatorial code designs for grant-free access in \ac{URLLC} we refer to~\cite{Boyd18}. 
Similarly, ~(\ref{es_prop_3}) provides connection to code constructions based on cyclic permutation matrices, providing compact protograph representations and offering insights to the decoding properties via the investigation of trapping sets (see \cite{Diao16} for a related discussion on the decoding properties of \ac{LDPC} codes).
%
%
In the following we describe the details of the existence - and construction of sparse signatures based on Euler squares and discuss the specific implications on the proposed coding scheme. 
\subsection{Graph Theoretical Background}
In this section we provide the necessary graph-theoretic preliminaries with focus on bipartite graphs and partial geometries. 
%
%
A graph $G\left(\mathcal{V}, \mathcal{E}\right)$  with vertex set $\mathcal{V}$ and edge set $\mathcal{E}$ is denoted as \textit{bipartite}, if its vertex set $\mathcal{V}$ can be expressed as the union of two sets $\mathcal{U}~\cup~\mathcal{W}$ such that all edges of $\mathcal{E}$ are between vertices in the sets $\mathcal{U}$ and $\mathcal{W}$.
A \textit{walk} of length $k$ in a graph $G$ is a sequence of vertices $(v_0, v_1, \ldots, v_k)$ such that any consecutive vertices  $(v_i,v_{i+1}),~\forall i \in \{0,1,\ldots, k-1\}$ form an edge in $\mathcal{E}$. 
The walk is of length $k$ if it traverses $k$ edges. 
A walk is closed, if $v_k = v_0$. 
Further, a walk is a \textit{cycle}, if the vertices of the walk are distinct (except for $v_k = v_0$) and cycle-free, if it is a closed walk with no cycles. 
%
%
The \emph{girth} 
of the graph is the length of the shortest cycle in $G$.
A graph $G$ with vertex set $\mathcal{V}$ can be represented as a $v \times v$ adjacency matrix $\boldsymbol{A}$ with $\left[\boldsymbol{A}\right]_{i,j} = 1$, if $(v_i,v_j)\in \mathcal{E}$ and zero otherwise, where $\vert \mathcal{V} \vert=v$.
%
%
The set of (real) eigenvalues of $\boldsymbol{A}$ is referred to as the spectrum of the graph. 
A $(k,n)$~-~bi-regular bipartite graph is a graph whose $u=\vert \mathcal{U} \vert$ left (or variable) nodes have degree $k$ (incident with $k$ edges), and whose $w=\vert \mathcal{W} \vert$ right (or check) nodes have degree $n$, while the number of edges is $uk=wn$. 
%
%
%
%
%
%
%
\subsection{Construction from Euler Squares}
\label{sec:euler_construction}
%
%
%
%
As discussed in the introduction of this chapter, an Euler square $E(\gamma, \rho)$ of order $\gamma$ and degree $\rho$ can be defined as a square array of $\gamma^2$~$\rho$-tuple of numbers~$\left(a_{ij1}, a_{ij2}, \ldots a_{ij\rho}\right)$, where $a_{ijr}\in\left\{0, 1, 2, \ldots, \gamma-1\right\}$ with $r=1, 2, \ldots, \rho$;~$i,j=1, 2, \ldots, \gamma$;~$\gamma>\rho$; $a_{ipr}\neq a_{iqr}$ and $a_{pjr}\neq a_{qjr}$ for $p\neq q$ and $a_{ijr} a_{ijs} \neq a_{pqr}a_{pqs}$ for $i\neq p$ and $j\neq q$.
Explicit constructions of Euler squares are known to exist for the following cases\cite{macneish1922euler}:
\begin{itemize}
\item $E(p, p-1)$, where $p$ is a prime number.
\item $E(p^r$, $p^r-1)$, where $p$ is a prime number.
\item $E(\gamma,\rho)$, where $\gamma=2^r p_1^{r_1}p_2^{r_2}\ldots p_l^{r_l}$ for distinct odd primes $p_1, p_2, \ldots, p_l$, and \\ ${\rho+1=\min \left\{2^r, p_1^{r_1}, p_2^{r_2}, \ldots , p_l^{r_l}\right\}}$.  
\end{itemize}
Furthermore, the existence of the Euler square $E(\gamma, \rho)$ implies that the Euler square $E(\gamma, \rho')$, with $\rho' < \rho$, also exists.
%
%
Based on the above, for $\gamma\geq 3$, $\rho\geq 2$, the binary matrix $\boldsymbol{F}$ of size $\gamma \rho \times \gamma^2 $ can be constructed as: 
\begin{align}
[\boldsymbol{F}]_{i,j}&=\left\{\begin{array}{ll}1 & \mathrm{if}\ {({a}_{j})}_{\lfloor\frac{i-1}{\gamma}\rfloor+1}\equiv (i-1) \mathrm{mod}\ \gamma\\ 
0 & \mathrm{otherwise}\end{array}\right\}, \label{eq:ES}
\end{align}
where $(a_j)$ is the $j$-th $\rho$-tuple, $(a_j)_l$ is the $l$-th element in the $j$-th $\rho$-tuple and $\lfloor x \rfloor$ denotes the largest integer not greater than $x$. 
Consequently, $\boldsymbol{F}$ is effectively a block matrix consisting of $\rho$ number of $\gamma\times \gamma^2$ blocks, where there are exactly $\rho$ ones in each column of $\boldsymbol{F}$ and each column of $\boldsymbol{F}$ correspond to a $\rho$-ad in the Euler square $E(\gamma,\rho)$.
Following \cite[Theorem 1]{Diao16}, we observe that $\boldsymbol{F}^T$ is the line-point incidence matrix of a partial geometry $\mathrm{PaG}(\gamma, \rho, \rho-1)$ with $n=\gamma\rho$ points corresponding to the columns of $\boldsymbol{F}^T$ and $m=\gamma^2$ lines corresponding to the rows of $\boldsymbol{F}^T$ (details on partial geometries are provided in  Appendix~\ref{sec:app_part_geom}).
Specifically, the associated partial geometry $\mathrm{PaG}(\gamma, \rho, \rho-1)$ based on an Euler square $E(\gamma, \rho)$ is \ac{QC}, due to the \ac{QC} structure of the associated line-point incidence matrix (see \cite{Diao16} for a general overview of quasi-cyclic partial geometries).
%
%
While the construction from the $E(\gamma, \rho)$ is a quasi-cyclic partial geometry $\mathrm{QC-PaG}(\gamma, \rho, \rho-1)$, not all partial geometries of the type $\mathrm{QC-PaG}(s, t, t-1)$ correspond to an Euler square construction, since Euler squares do not exist for all $s,t>1$. 
%
%

Further, a $\mathrm{PaG}(\gamma, \rho, \rho-1)$ can be represented as protograph with $\gamma$ (super) \acp{VN} and $\rho$ (super) \acp{CN}, that contains all the structural information of the matrix $\boldsymbol{F}$ derived from $E(\gamma, \rho)$ (see Appendix~\ref{sec:app_protograph} for details).
Hence, the construction possess simple representations to realize efficient decoding that allows for high-speed iterative decoding implementation using belief propagation~\cite{Fang2015, Divsalar2009,Abbasfar2007}.
%
%

%
%
%
\subsection{Implications on Encoding/Decoding}
\label{sec:implications_decoding}
In this section we describe implications which result from the specific properties of the sparse signature construction based on Euler squares in combination with the proposed coding scheme. 
\subsubsection{Peeling Decoder}
\begin{figure}[t!]
\centering
\begin{subfigure}[c]{0.24\columnwidth}
\includegraphics[width=\columnwidth]{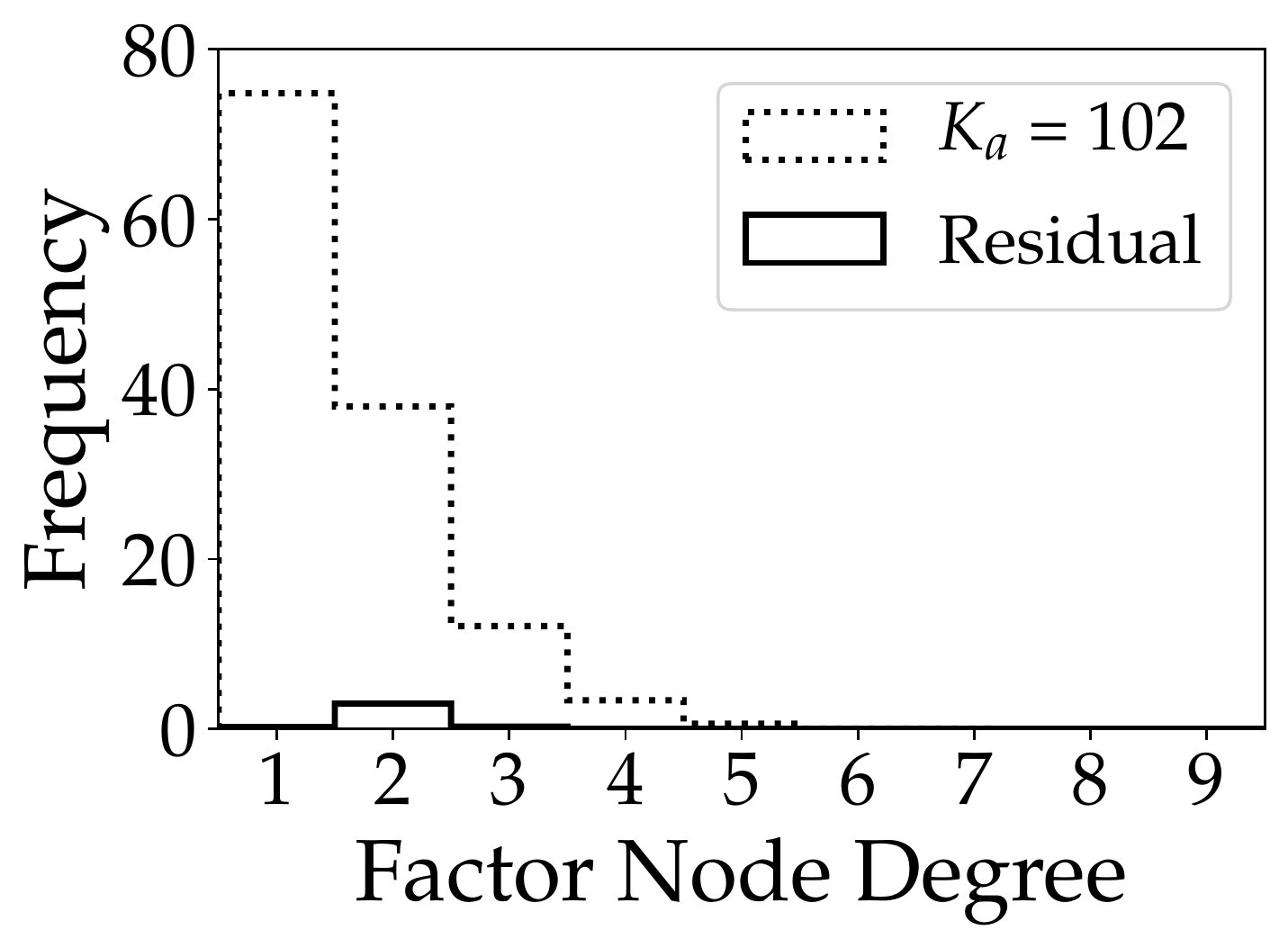}
\subcaption[]{$\alpha = 0.009$}
\end{subfigure}
\begin{subfigure}[c]{0.24\columnwidth}
\includegraphics[width=\columnwidth]{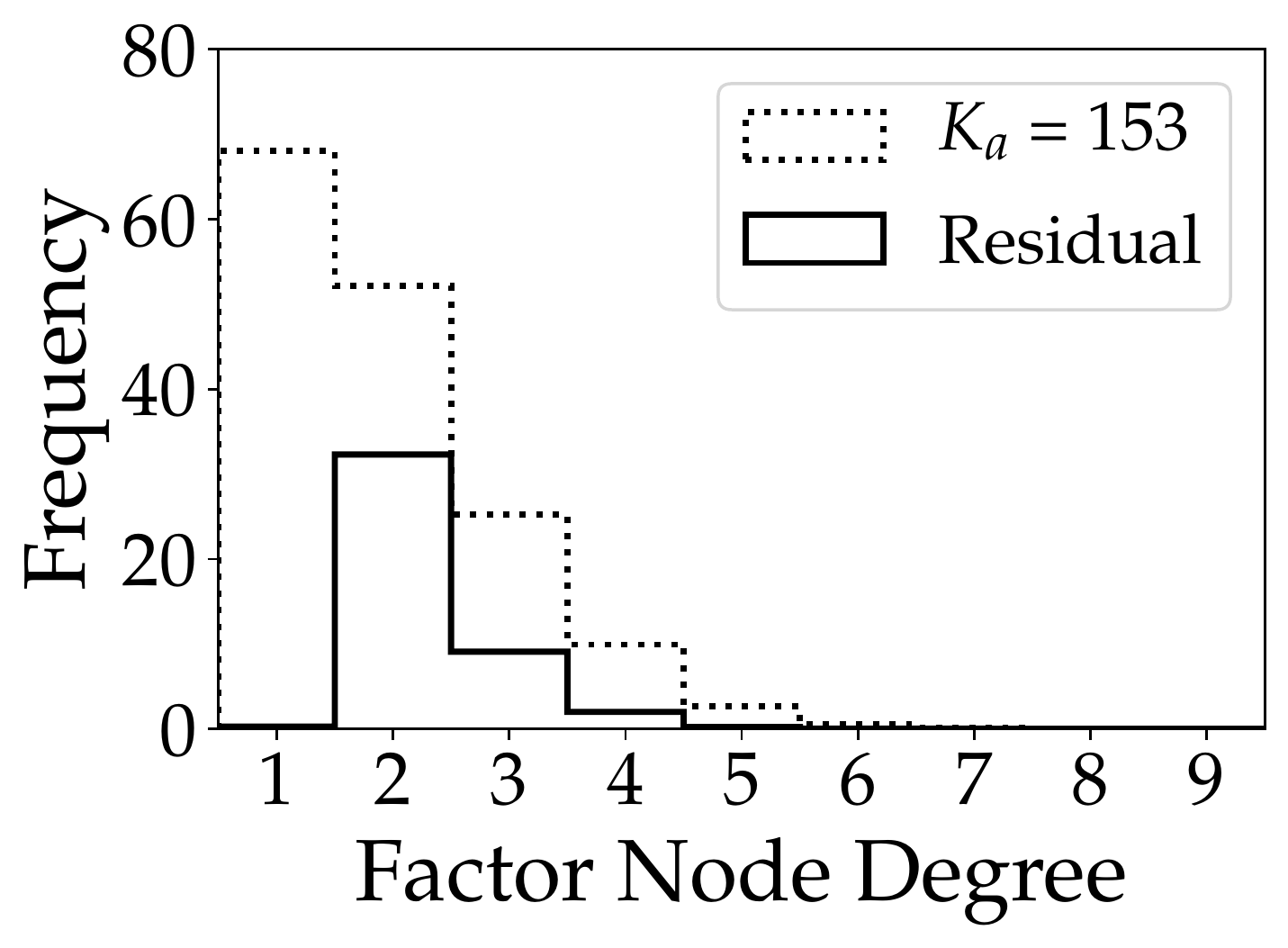}
\subcaption[]{$\alpha = 0.014$}
\end{subfigure}
\begin{subfigure}[c]{0.24\columnwidth}
\includegraphics[width=\columnwidth]{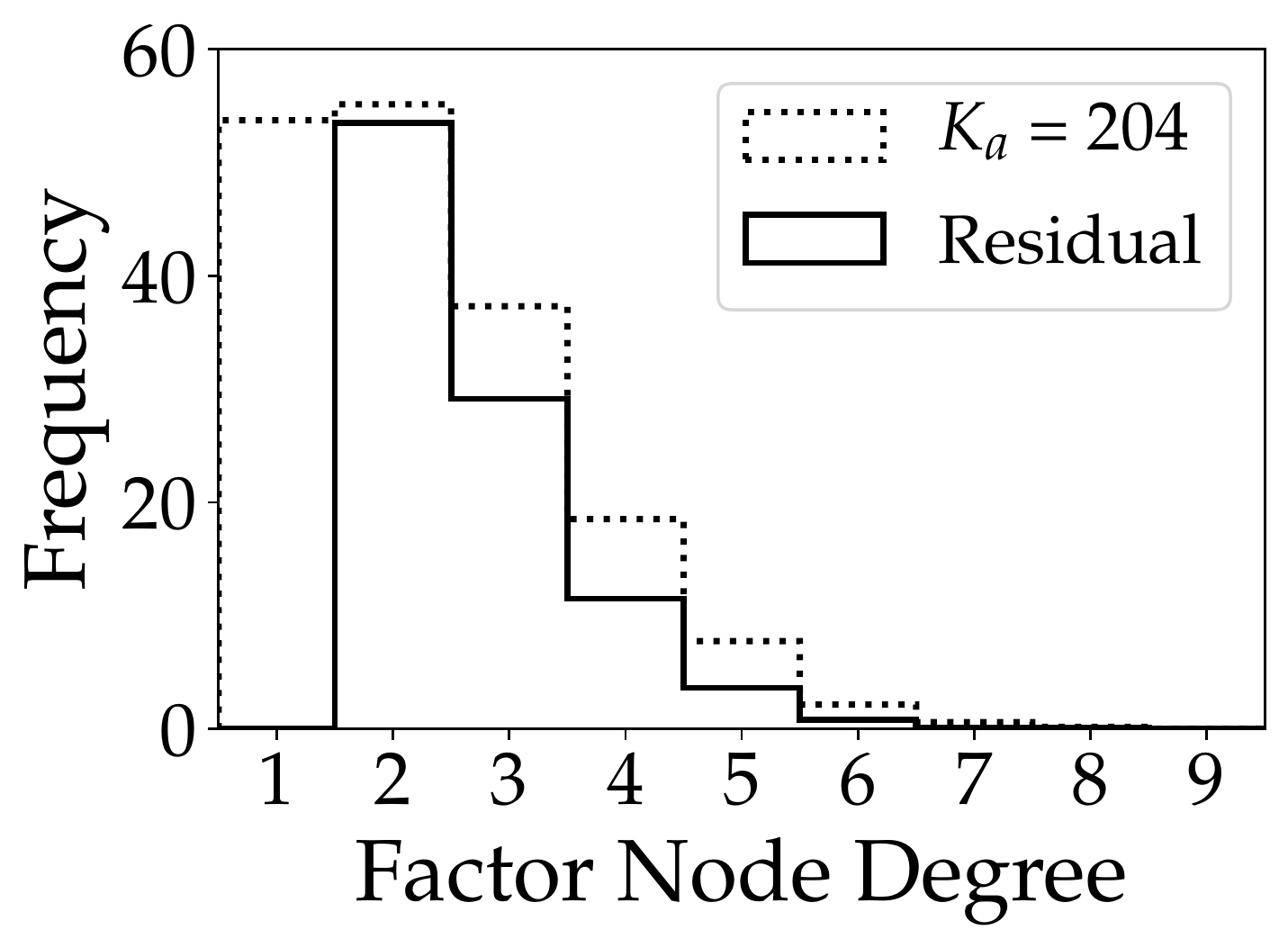}
\subcaption[]{$\alpha = 0.018$}
\end{subfigure}
\begin{subfigure}[c]{0.24\columnwidth}
\includegraphics[width=\columnwidth]{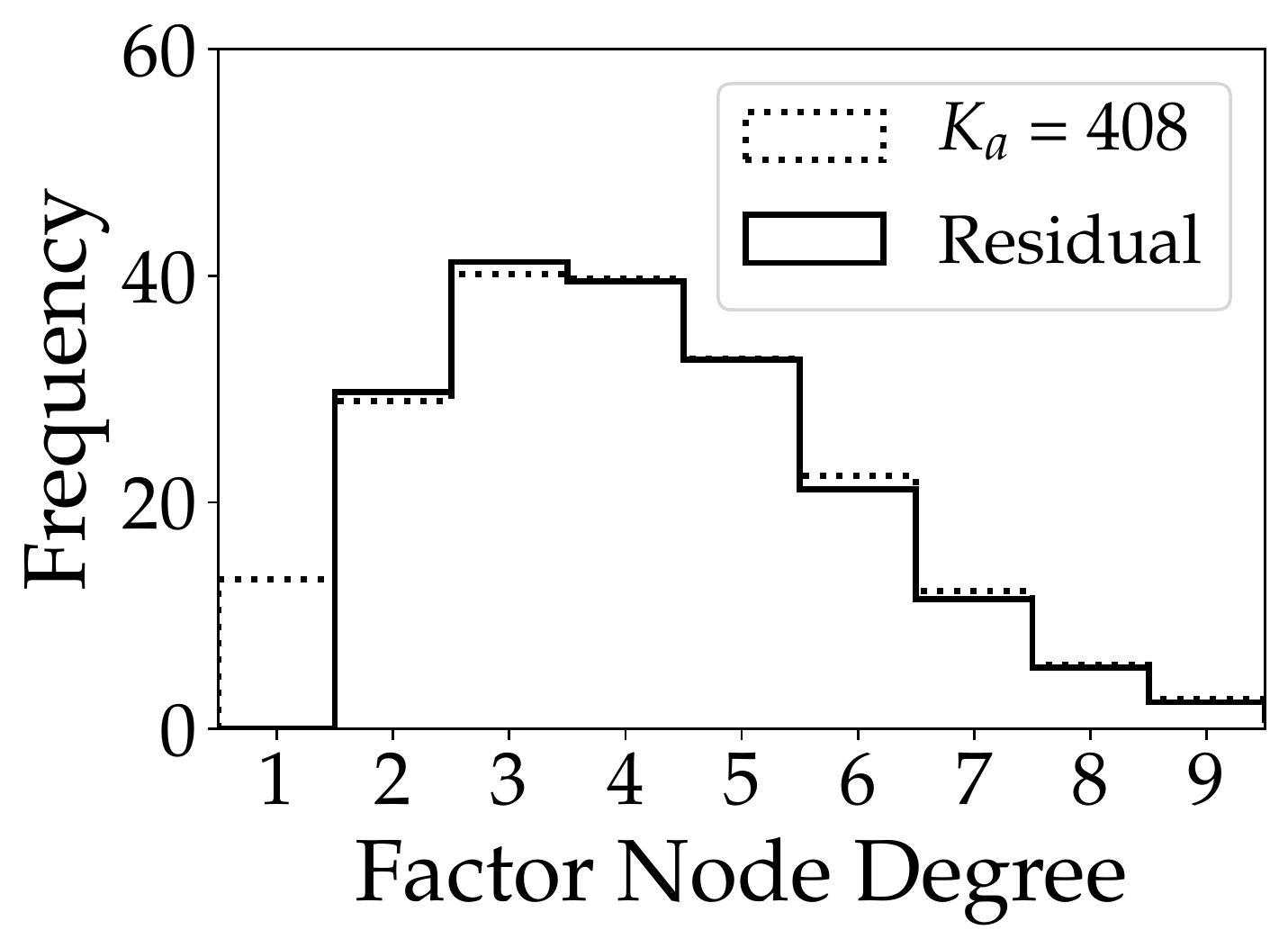}
\subcaption[]{$\alpha = 0.036$}
\end{subfigure}
\caption{Empirical factor node degree distribution before (dashed) and after (solid) peeling-decoding for the bipartite graph based on $E(101, 2)$ at different numbers of active devices $K_a$.}
\label{fig:PeelingFNDegreeDistribution}
\end{figure}

%
The analysis in \cite{zeng2016peeling} on the asymptotic behavior of peeling decoding for \ac{LDPC} codes shows that there is a threshold value based on the density factor $\alpha = K_a / K$, if below this threshold, the recovery algorithm is successful, otherwise it will fail whereas the threshold is dependent on the parameter of the sensing matrix.
In contrast, we deploy the peeling decoder to successively decode and remove messages block-wise from the received signal (as discussed in Section~\ref{sec:Receiver_Processing}) for the case of sparse activation ($\alpha < 1$).  
By design, the proposed signature construction guarantee that any two rows (columns) of $\boldsymbol{F}$ have at most one place where both have non-zero entries, i.e. there is an overlap in one position at most. 
For the peeling decoder, this property ensures that no length-4 cycle, i.e. two variable nodes are connected to the same factor node, exists within any randomly (pruned) factor graph corresponding to $\boldsymbol{F}(\mathcal{K}_a)$ for any $\mathcal{K}_a \subset \mathcal{K}$.
The impact of the density ($\alpha$) on the performance of the peeling decoder is depicted in Fig.~\ref{fig:PeelingFNDegreeDistribution}, where the empirical factor-node degree distribution of the pruned (dashed line) - and the residual factor graph after peeling decoding (solid line) is depicted for a construction based on $E(101,2)$ for different sparsity level $K_a$. 
We note that the sparse construction based on Euler-square $E(101, 2)$ supports in total $K = 11~201$ devices 
with signature length $n_s = 202$ and $2$ non-zero elements per signature. 
%
The large number of signatures $K$ relative to the signature length $n_s$ ensures that even at very low activity $\alpha$, the number of active devices is still reasonable at $K_a = \alpha K$.
%
From Fig.~\ref{fig:PeelingFNDegreeDistribution}, it can be observed that with $K_a = 102$ active devices (messages) almost $2/3$ of all factor nodes of the pruned factor graph (dashed line) are of degree $1$ and that almost all nodes can be decoded and that the residual factor graph after peeling decoding (solid line) contains only a small number of factor nodes of degree~$2$ and~$3$.  
It can be further observed, that by increasing the number of active devices (messages), the number of degree $1$ factor nodes, and consequently the amount of decodable messages for the peeling process decreases. For example with $K_a = 408$, only $12$ nodes are of degree-$1$ and after removing these nodes, the residual factor graph (solid line) remains almost similar.  
%

%

\subsubsection{Message Passing}
For a sparse signature design associated with a bipartite graph $G(\mathcal{V}, \mathcal{C}, \mathcal{E})$, its error performance depends on a number of structural properties of $G$. One of these properties is the girth, which is the length of the shortest cycle in $G$. For the bipartite graphs associated with the partial geometries $\mathrm{PaG}(\gamma, \rho, \rho-1)$ considered here, it has been shown in \cite{Diao16} that the girth is $8$ when $\rho=2$, and there are $\frac{1}{4}\gamma^2(\gamma-1)^2$~cycles of length $8$~\cite{Johnson2004}. 
%
%
%
Further, for $\rho>2$, the girth of the associated graph is $6$ and there are $\frac{1}{6}\gamma^3(\gamma-1)(\rho-2)(\rho-1)$
%
%
%
cycles of length $6$ \cite{Johnson2004}. Besides the impact on the decoding procedure, the size of the girth also provides a hint on the number of iterations that need to be performed in each \ac{MPA} step of the turbo-decoding procedure. 
%
%
Another important property of $G$ is \textit{connectivity}, defined
as the number of \acp{VN} which are connected to a specific \ac{VN} by paths of length $2$, which has an effect on the rate of convergence of the message passing procedure. For the $(\gamma, \rho)$-biregular graphs of interest here, the connectivity is $\rho (\gamma-1)$. In general, the message passing algorithm running on a bipartite graph with higher connectivity converges faster.
Besides the girth and the connectivity, an important factor is also the size of the so called \textit{trapping sets}. The analysis of trapping sets for the graphs of interest here is beyond the scope of the paper. We note that the problem of determining the sizes of trapping sets for \ac{LDPC} codes obtained from finite geometries was initiated in \cite{Johnson2004}. In the special case of partial geometries, some bounds were presented in \cite{Diao16}. 
\section{Performance Evaluation}
\label{sec:Results}
The proposed coding scheme based regular sparse constructions derived from  $E(\gamma, \rho)$, as discussed in Section~\ref{sec:code_design}, is flexible in the sense that it can be explicitly configured for a wide number of system parameters, i.e. total number of users / messages $K = \gamma^2$; signature length $n_s = \gamma \rho$; number of non-zero elements (per signature) $\rho$; and \ac{FEC} blocklength $\ell = N / n_s$, where $N$ denotes the overall number of channel uses. 
\begin{Remark}
We note that for grant-based (scheduled) case, conventionally the \textit{system load} $\beta$ is defined in the literature on \ac{NOMA} as the average number of devices per resource. According to our coding scheme this is given as $\beta = \gamma / \rho$.
For grant-free case (random access), typically the number active ($K_a$) users / messages are of relevance. 
\end{Remark}
\subsection{Spectral Efficiency with Optimum Decoding}
\begin{figure}[hbtp]
\centering
\includegraphics[width=0.5\columnwidth]{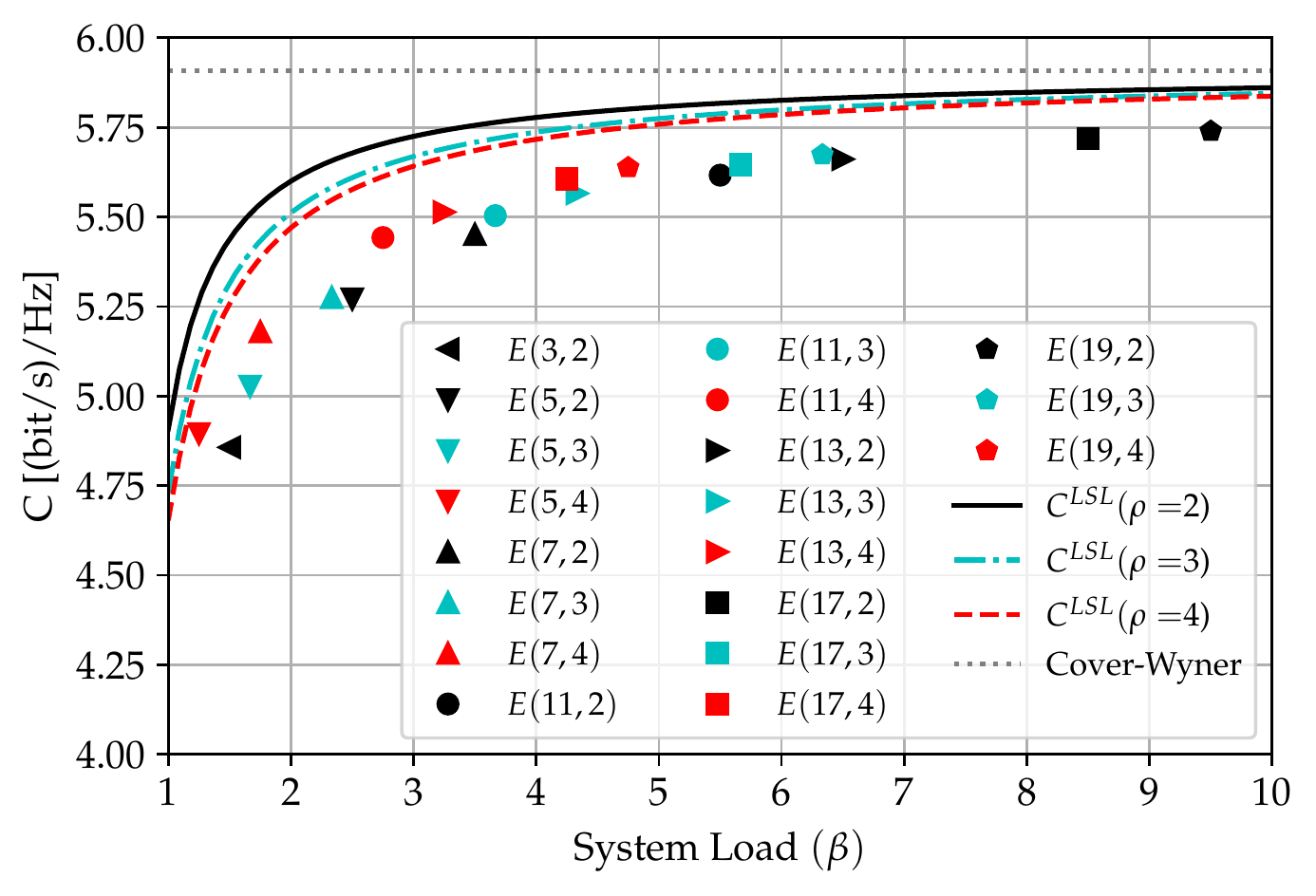}
\caption{Normalized achievable throughput for constructions based on $E(\gamma, \rho)$ at ${E_b/N_0 = 10~\text{dB}}$, where $\rho$ is the number of non-zero elements, $\gamma \rho$ the signature length and $\gamma^2$ the total number of signatures, compared to~\cite{Zaidel2018}, denoted as $C^{LSL}$.}
\label{fig:NumResults_10dB}
\end{figure}

We evaluate the spectral efficiency of our proposed regular-sparse signature constructions with finite signature length ($n_s$) in terms of the maximum achievable throughput per resource element, which holds for infinite \ac{FEC} blocklength ($\ell \rightarrow \infty$), defined as 
\begin{align}
C^{Opt}_{n_s}(\boldsymbol{S}) &= \frac{1}{n_s} \sum_{i=1}^{n_s} \text{log}_2\left(1 + \text{SNR}\cdot \lambda_i\left(\boldsymbol{S}\boldsymbol{S}^H\right)\right)\label{eq:spectral_efficiency},
\end{align}
where $\lambda_n(\boldsymbol{X})$ corresponds to the $n^{\text{th}}$ eigenvalue of $\boldsymbol{X}$. 
In Fig.~\ref{fig:NumResults_10dB}, we compare the spectral efficiency of the proposed constructions based on Euler squares against the asymptotic performance of regular-sparse signatures in the \ac{LSL} ($n_s \rightarrow \infty$) from~\cite{Zaidel2018} and the ultimate Cover-Wyner upper bound for overloaded systems\cite{verdu1999spectral}, which corresponds to the absence of spreading. 
We observe that the constructions from Euler-squares, although having finite signature size, perform fairly well across the entire system load range $\beta = K/n_s$. 
For example, the solid black line in Fir.~\ref{fig:NumResults_10dB} corresponds to the limiting spectral efficiency of regular sparse spreading with $2$ non-zero elements in the \ac{LSL}. Accordingly, the black marker correspond to finite-length constructions with $2$ non-zero elements based on Euler squares, i.e. $E(\gamma, \rho = 2)$. We observe that the gap to the \ac{LSL} decreases with increasing signature length $n_s = \rho\gamma$. 
\subsection{Grant-based Access with Finite-size FEC}
\begin{figure}[hbtp]
\centering
\includegraphics[width=0.5\columnwidth]{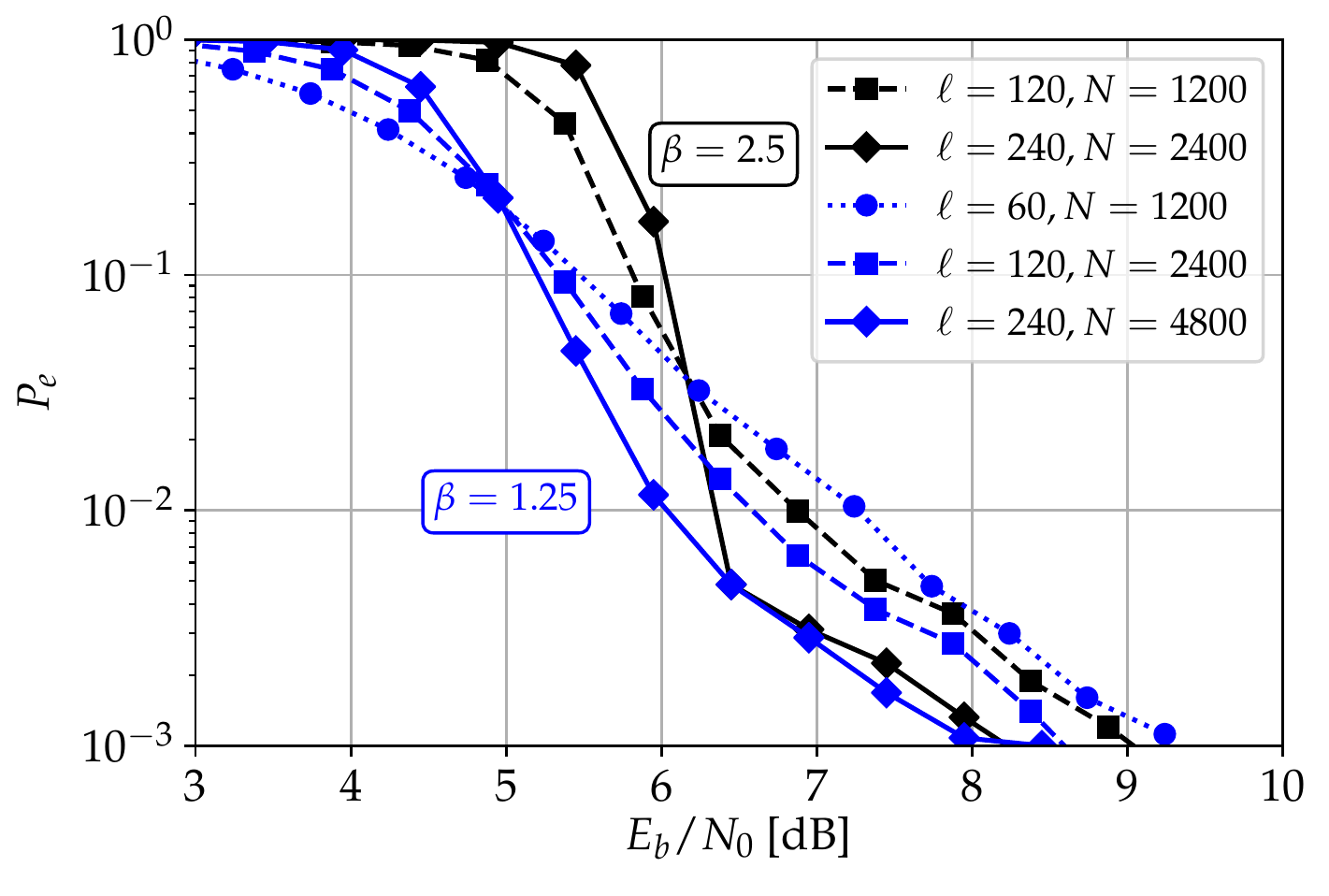}
\caption{Error vs $E_b/ N_0$ for different code constructions and \ac{FEC} blocklength.}
\label{fig:Pe_vs_SNR}
\end{figure}

\begin{figure}[hbtp]
\centering
\includegraphics[width=0.48\columnwidth]{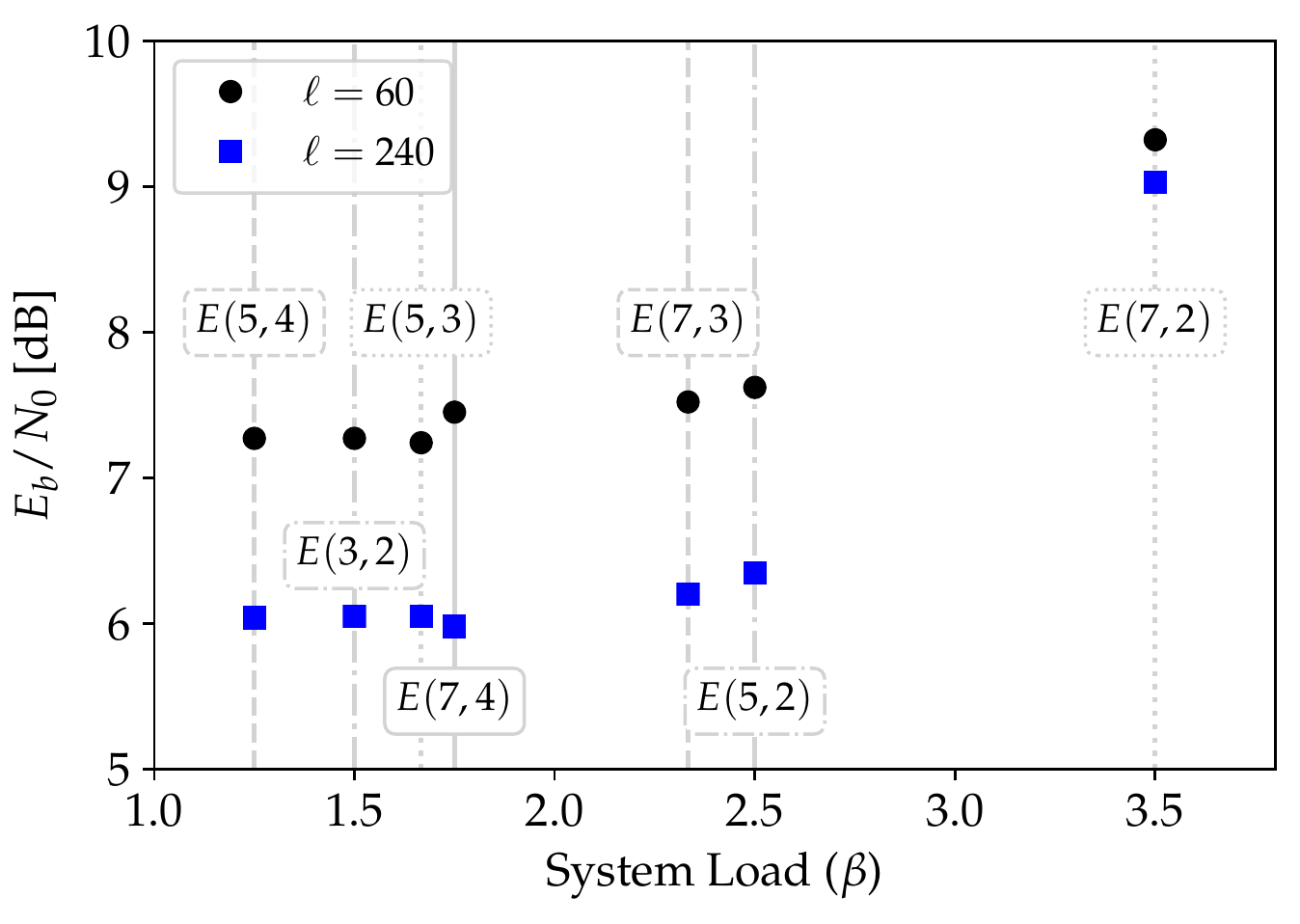}
\caption{Required $E_b / N_0$ to achieve $P_e \leq 0.05$ as a function of the system load ($\beta$).} 
\label{fig:EbN0_vs_Overload_n60}
\end{figure}

%
We consider a non-orthogonal multiple access scenario, where users are co-scheduled on the same physical resources, where the activity pattern as well as the seed of the \ac{FEC} per device can be assumed to be known at the receiver. 
In this context, we are specifically interested in characterizing the system in the \ac{FBL} regime, i.e. $N = n_s \cdot \ell \ll \infty$. 
The figure of merit is the average error probability, defined as
\begin{align}
    P_e  & = \frac{1}{K} \sum_{k \in \mathcal{K}} \text{Pr}\left\{w_k \neq \hat{w}_k~\vert~w_k\right\}. \label{eq:P_e}
\end{align}
The encoding is done as described in Section~\ref{sec:coding_scheme} and 
throughout the following if not stated otherwise, 
we consider a regular \ac{LDPC} code as \ac{FEC} with parity-check matrix according to Gallager's algorithm~\cite{GallagerLDPC} and $R_{FEC} = 0.5$. 
The trade-off between the channel coding gain (as a function of the \ac{FEC} blocklength $\ell$) and the system load as a property of the specific Euler square construction is illustrated in  Fig.~\ref{fig:Pe_vs_SNR}, where the error rate performance for different \ac{FEC} block lengths ${\ell = \{60, 120, 240\}}$ is plotted. 
%
%
From Fig.~\ref{fig:Pe_vs_SNR}, one can observe the following: 
(\emph{i}) a larger message size improves the performance in terms of $P_{e}$ in the lower range of the \textit{water-fall} region, for both constructions. This effect is directly connected to the higher coding gain which comes with the larger \ac{FEC} blocklength;
(\emph{ii}) for a fixed number of resources ($N$), the system operated at higher overload, i.e. $E(5,2)$, requires a higher energy-per-bit to achieve a similar performance compared to $E(5,4)$. 
Fig.~\ref{fig:EbN0_vs_Overload_n60} plots the required energy-per-bit $E_b / N_0$ to achieve a target error of $P_e \leq 0.05$ as a function of the overload, which is achieved by using different Euler-square constructions. 
From Fig.~\ref{fig:EbN0_vs_Overload_n60}, we can observe two effects: (\emph{i}) the cost in terms of energy-per-bit only increases moderately with the overload in the observed range and (\emph{ii}) that a larger message size improves the energy efficiency due to the higher coding gain of the \ac{FEC}.
For example, at blocklength $\ell = 60$, the required energy-per-bit increases from $7.3~\text{dB}$ at $\beta = 1.25$ to $7.6~\text{dB}$ at $\beta = 2.5$ to achieve a similar performance in terms of $P_e$. 
Further, the higher coding efficiency, as an effect of a larger \ac{FEC} blocklength, increases the energy efficiency by $\approx 0.6~\text{dB}$ when the blocklength is increase by a factor of $2$.
%
%
%
%
\subsection{Grant-free Massive Access with Random User Activation}
\begin{figure}[thp]
\centering
\includegraphics[width=0.5\columnwidth]{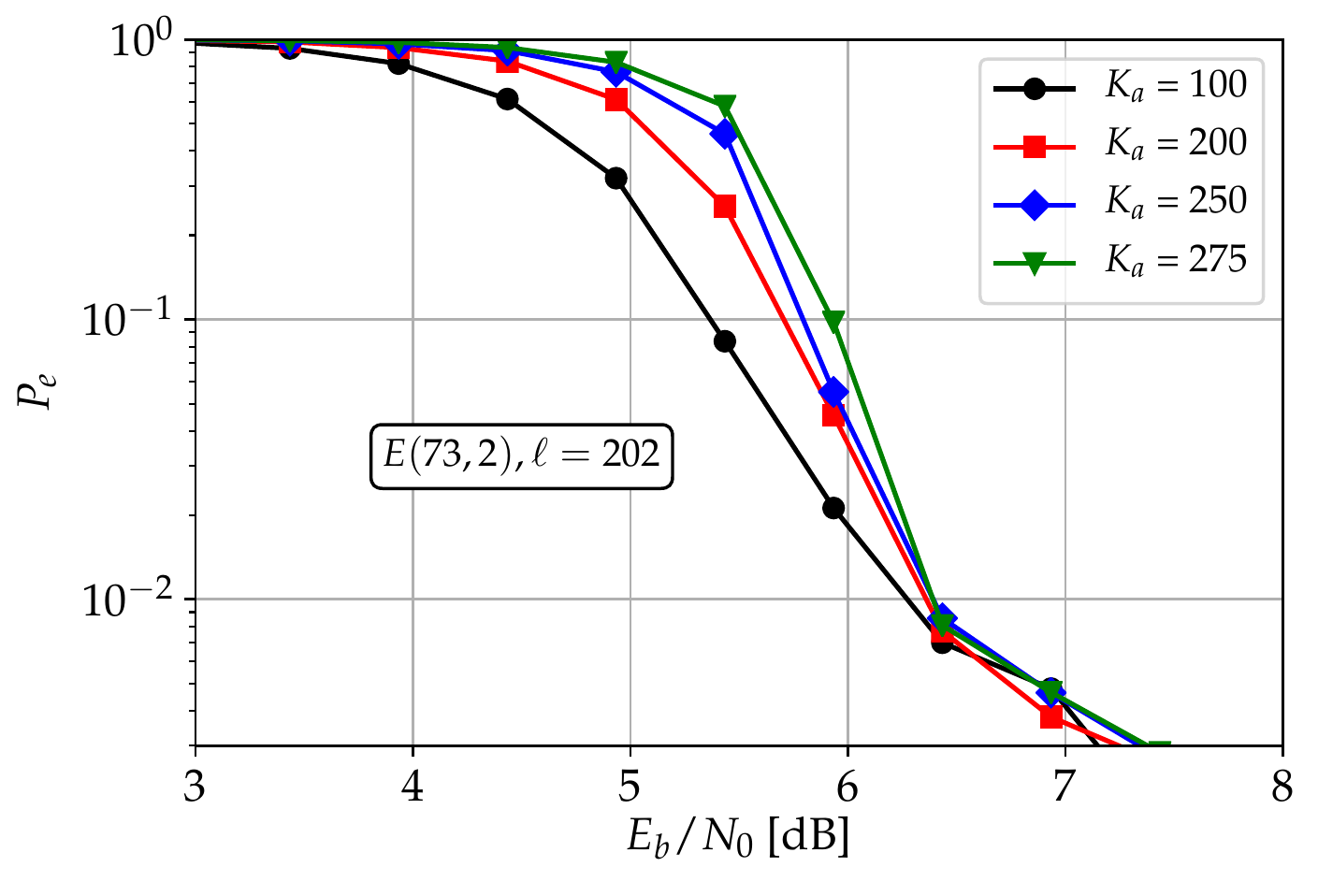}
\caption{Error-rate ($P_e$) as a function of the $E_b/ N_0$ for different number of active devices $K_a$ and  payload size $k = 101~\text{bit}$ and in total $N = 28~908$ shared resources. The code is based on $E(73,2)$, which supports in total $K = 5329$ devices.}
\label{fig:PEvsKa}
\end{figure}

Consider a (massive) grant-free random access protocol, where only a small subset $\mathcal{K}_a \subset \mathcal{K}$ with $K_a/K \ll 1$ are active simultaneously and the activity pattern is unknown to the receiver. 
In this regime, typically the total number of devices $K$ connected to the network is in the order of the number of available resources $N$ (or higher). 
Let $\mathcal{K}^t_a \subset \mathcal{K}$ be a random subset of active devices at slot $t$. %
With $\vert\mathcal{K}^t_a\vert = K_a$, we refine the error probability~\eqref{eq:P_e} as
\begin{align}
    P_e  & = \lim\limits_{T \rightarrow \infty}{\frac{1}{T\cdot K_a} \sum_{t \in [T]} \sum_{k \in \mathcal{K}^t_a} \text{Pr}\left\{w_k \neq \hat{w}_k~\vert~w_k\right\}}.
\end{align}
For the numerical simulations we will assume that a sufficiently large fraction of the available resources is reserved for the estimation of the active set (i.e. the pruned graph structure), such that reliability of this estimation step meets a required level.      
%
In Fig.~\ref{fig:PEvsKa}, the error rate $P_e$ is plotted as a function of $E_b/ N_0$ for different number of active devices $K_a$. 
In this setting, the coding scheme is $E(73,2)$ with \ac{FEC} with $\ell = 202$ payload size $k = 101$).
Based on this configuration, in total $K = 5329$ devices are supported at a total number of $N = 28~908$ resources.
We observe a moderate increase in the required energy-per-bit of $0.6~\text{dB}$ to achieve an error $P_e \leq 0.05$ when the number of active devices increases from $K_a = 100$ to $K_a = 275$. Notably, for all activity values $K_a$ at \ac{SNR} $>6.5~\text{dB}$, the characteristic error-floor of the \ac{FEC} appears at $P_e \le 10^{-3}$.
\subsection{Evaluation in the Context of Unsourced Random Access}
\begin{figure}[bthp]
\centering
\includegraphics[width=0.5\columnwidth]{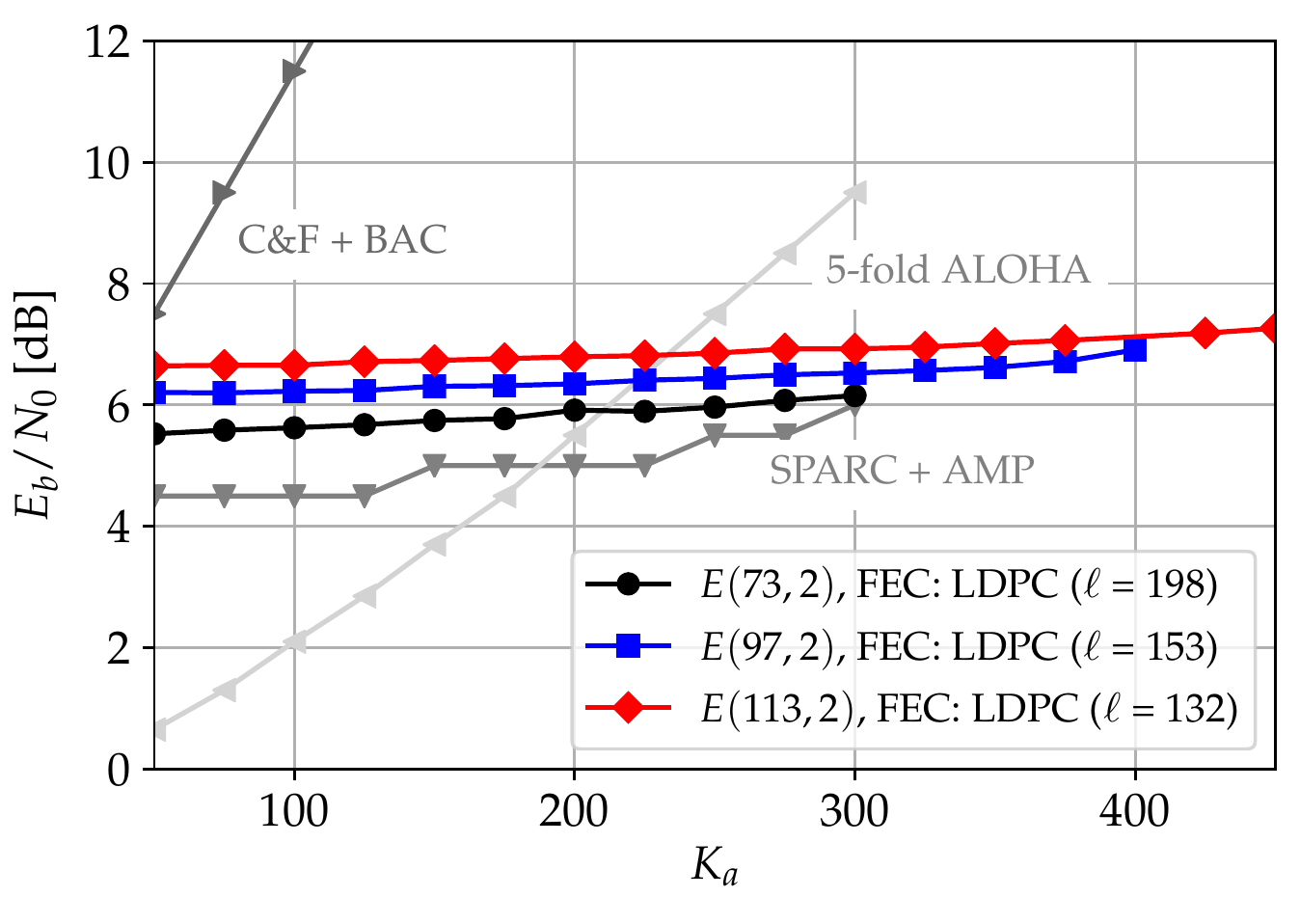}
\caption{Required $E_b / N_0$ at target error probability ${P_e \leq 0.05}$ as a function of the number of active devices $K_a$, for different code configurations at payload $k = 100~\text{bit}$.}
\label{fig:EbN0_vs_Ka}
\end{figure}

In the context of the \ac{U-RA} model~\cite{polyanskiy17}, our scheme can be modified (see Remark~\ref{remark:ura}) by letting the active users choose randomly a sparse sequence from a shared codebook (i.e. a set of sparse signatures obtained from an Euler square construction) and an associated interleaver pattern, in combination with a finite (short) blocklength \ac{FEC} code. 
Different to other \ac{U-RA} approaches, in our setting the joint \ac{MPA} and~\ac{FEC} decoding at the receiver decouples the received signal into parallel channels, yielding a decoding performance that is essentially determined by the structure of the (finite-length)~\ac{FEC} code.
In Fig.~\ref{fig:EbN0_vs_Ka}, we evaluate the performance of our scheme in the \ac{U-RA} setting (in the context of the mentioned recent works). As in \cite{ordentlich2017low} (denoted as "C\&F+BAC") and \cite{Fengler2019} (denoted as "SPARC+AMP"), we plot the $E_b / N_0$ (as a function of the the number of active devices $K_a$) required to achieve a target error probability $P_e \leq 0.05$ with a (per device) message size of $k\approx100$ bits, and a total number of  $N\approx 30~000$ resources shared by all $K$ devices. 
In this setting, we consider different code configurations with sparse signatures constructed from the Euler-square designs $E(73,2)$, $E(97,2)$ and $E(113,2)$ respectively and adjusted \ac{FEC} code rate accordingly. 
The results show that the required energy per bit (per device) increases by $\approx 0.7~\text{dB}$ for the configuration with $E(73,2)$ and $\ell = 198$, when the number of active devices increases from $50$ to $300$. 
The configuration using signatures $E(97,2)$ and $\ell = 153$, on the other hand, requires only $\approx 0.2~\text{dB}$ increase in the required energy. 
Importantly, the results indicate that the signature size can be traded with the \ac{FEC} coding rate to optimize the performance depending on the system load (the number of active users in the system), especially when the number of active devices exceeds $K_a \geq 300$, where recent \ac{U-RA} schemes show a rapid increase in  required energy \cite{Pradhan2019, Marshakov2019, Amalladinne2020, Fengler2019}.
\section{Discussion and Future Work}
\label{sec:Summary}
%
The presented simulation results illustrate the potential of the proposed approach as a viable solution for massive non-orthogonal access, both for scheduled (i.e. grant based) and grant-free transmissions. 
Here, key roles play (\emph{i}) the regular sparse signature design (i.e. the structure of the induced bipartite graph on which the message passing procedure for user separation takes place) and (\emph{ii}) the finite blocklength \ac{FEC} code applied on the level of individual user. 
For both aspects, the proposed coding scheme provides the flexibility to trade relevant system parameters like number of users, available resources via the channel coding rate, as shown in Fig.~\ref{fig:Pe_vs_SNR}~\&~\ref{fig:EbN0_vs_Overload_n60}.

In a direct comparison with apparently conceptually similar schemes such as \ac{SCMA}, we argue that in the massive access scenario, it is worth investing more effort in the sparse mapping rather than in the multi-dimensional constellation optimization, which quickly becomes computational prohibitive as soon as the number of devices grow. %
In contrast to \ac{SCMA}, our design is easily scalable to a large number of users and shared resources. 
In addition, we evaluate performance in the event of random user activation, which is not taken into account in SCMA (at least not in the context of “massive IoT”). 
What is important here is the sparse signature design from the receiver's point of view, which is now performed on a pruned- instead of a full factor graph. 
Hence, a direct comparison with “conventional SCMA” in the massive connectivity scenario is difficult due to the complexity of the SCMA codebook design. 
Therefore, we focus on the investigation of the interplay of the different system parameters in our scenario.

We observed that the $E_b/{N_0}$ performance of the proposed construction is quite robust to the increase of the system load $\beta$ in the case of scheduled transmissions, respectively to the number of active users $K_a$ in the grant-free random access setting, as illustrated by Fig.~\ref{fig:EbN0_vs_Overload_n60} and Fig.~\ref{fig:PEvsKa},  respectively.    %
In that context further investigation is needed for the overall code design and parameterization, e.g. the \ac{S-EXIT} chart framework\cite{Ebada2018} might be employed to jointly optimize the parameterization of the sparse signatures and the \ac{FEC} code.
%
For \ac{U-RA}, we have modified our scheme by letting the active users choose randomly a sparse sequence from a shared set of sparse signatures, together with an associated interleaver pattern. The performed numerical simulations in Fig.~\ref{fig:EbN0_vs_Ka} illustrate the potential (in terms of energy-efficiency) of the proposed scheme in the scenario with higher user activation (e.g. beyond $K_a=300$ active users in the setting where the system users send fixed messages of $k=100$ bits over $N=30~000$ channel resources), compared to other approaches that show a steep  increase in the required $E_b/{N_0}$ in this regime. 
We note that in contrast to \cite{polyanskiy17}, the number of active devices $K_a$ does not need to be known perfectly in our scheme, hence enabling \emph{true} random access for \ac{U-RA}.
While here we have resorted to a particular \ac{LDPC} code construction for \ac{FEC}, for future work it would be interesting to consider other  finite-blocklength codes for \ac{FEC}, including polar and algebraic codes. 
In addition, it would be of interest to investigate the performance of the scheme in the presence of receive diversity, both  in a multi-antenna receiver scenario and in a multi-cell Cloud/Fog-Radio Access Network architecture. 
%
\appendix
\subsection{Partial Geometries}
\label{sec:app_part_geom}
Consider a system composed of a set of points $\mathcal{N}$ and a set of lines $\mathcal{M}$ (in which a line is defined as a set or points).
If a line $l \in \mathcal{M}$ contains a point $n \in \mathcal{N}$, we say that $n$ is on $l$ and $l$ passes through $n$.
If two points are on a line, the two points are adjacent and if two lines pass through the same point these two lines intersect (otherwise they are parallel). 
The system composed of the sets $\mathcal{N}$ and $\mathcal{M}$ is a \textit{partial geometry} $PaG(\gamma, \rho, \delta)$~\cite{Diao16}, if the following conditions are satisfied for some fixed integers $\gamma\geq 2$, $\rho\geq 2$ and $\delta\ge 1$: i) any two points are on one line; ii) each point is on $\gamma$ lines; iii) each line passes through $\rho$ points; and iv) if a point $n$ is not on line $l$, there are exactly $\delta$ lines, each passing through $v$ and a point on $l$.
%
%
%
%
%
%
%
%
%
%
%
The binary sparse matrix $\boldsymbol{F}^T$ which is derived from the Euler square $E(\gamma, \rho)$ constitutes the line-point incidence matrix of a partial geometry $\mathrm{PaG}(\gamma, \rho, \rho-1)$ with $n=\gamma\rho$ points corresponding to the columns of $\boldsymbol{F}^T$ and $m=\gamma^2$ lines corresponding to the rows of $\boldsymbol{F}^T$.
%
The associated bipartite graph 
%
%
%
has $m=\gamma^2$ \acp{VN} of degree $\rho$ (representing the $m$ lines in $\mathrm{PaG}(\gamma, \rho, \rho-1)$), and $n=\gamma\rho$ \acp{CN} of degree $\gamma$  (representing the $n$ points in $\mathrm{PaG}(\gamma, \rho, \rho-1)$). 
The associated partial geometry $\mathrm{PaG}(\gamma, \rho, \rho-1)$ is \ac{QC}, due to 
structure of the associated line-point incidence matrix. 
%
%
%
\subsection{Protograph Representation}
\label{sec:app_protograph}
\begin{figure}[hbtp]
\centering
\includegraphics[trim= 10 20 10 30,clip, width=0.6\columnwidth]{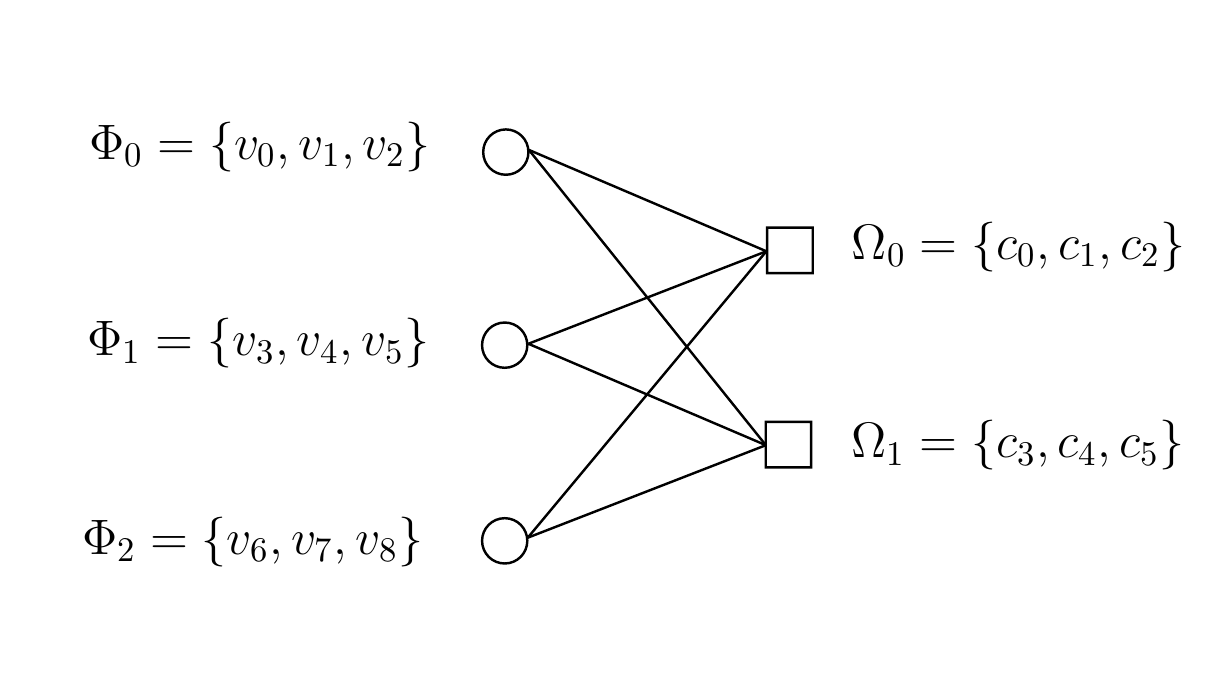}
\caption{Protograph representation of the bipartite graph associated with $E(3,2)$.}
\label{fig:ProtoExample2}
\end{figure}

The matrix $\boldsymbol{F}^T$ constructed from $E(\gamma, \rho)$  constitutes a binary $\gamma\times \rho$ array of \acp{CPM} of order $\gamma$ of the following form:
{\small
\begin{align} 
\boldsymbol{F}^T & = \begin{bmatrix} {\boldsymbol{C}}_{0,0} & {\boldsymbol{C}} _{0,1} & \cdots & {\boldsymbol{C}} _{0,\rho -1} \\ {\boldsymbol{C}}_{1,0} & {\boldsymbol{C}} _{1,1} & \cdots & {\boldsymbol{C}} _{1,\rho -1} \\ \vdots & \vdots & \ddots & \vdots \\ {\boldsymbol{C}}_{\gamma -1,0} & {\boldsymbol{C}} _{\gamma -1,1} & \cdots & {\boldsymbol{C}} _{\gamma -1,\rho -1}\\ \end{bmatrix}, 
\end{align}  
}
%
%
where $\boldsymbol{C}_{i,j}$ is uniquely specified by the location of the single $1$-entry of its top row, called the \textit{generator}. 
If the single $1$-entry of the top row of $\boldsymbol{C}_{i,j}$ is located at the position $k_{i,j}$, $0\leq k_{i,j}<\gamma$, then we use $(k_{i,j})$ to specify the \acp{CPM} $\boldsymbol{C}_{i,j}$. 
Following \cite{Diao16}, one can divide the \acp{VN} of $\mathrm{PaG}(\gamma, \rho, \rho-1)$ in $\gamma$ disjoint clusters $\Phi_0,\Phi_1,\ldots,\Phi_{\gamma-1}$, and the \acp{CN} in $\rho$ disjoint clusters, $\Omega_0, \Omega_1,\ldots, \Omega_{\rho-1}$ \acp{CN}.
Consequently, one can construct a (bipartite) \textit{protograph} with $\gamma$ (super) \acp{VN} and $\rho$ (super) \acp{CN}. The protograph $\mathrm{PaG}(\gamma, \rho, \rho-1)$ contains all the structural information of the matrix $\boldsymbol{F}$ derived from $E(\gamma, \rho)$. An example of a protograph associated with the construction from the Euler square $E(3,2)$ is depicted in Fig.~\ref{fig:ProtoExample2}.  
%
\bibliography{IEEEabrv, library}

\begin{thebibliography}{10}
\providecommand{\url}[1]{#1}
\csname url@samestyle\endcsname
\providecommand{\newblock}{\relax}
\providecommand{\bibinfo}[2]{#2}
\providecommand{\BIBentrySTDinterwordspacing}{\spaceskip=0pt\relax}
\providecommand{\BIBentryALTinterwordstretchfactor}{4}
\providecommand{\BIBentryALTinterwordspacing}{\spaceskip=\fontdimen2\font plus
\BIBentryALTinterwordstretchfactor\fontdimen3\font minus
  \fontdimen4\font\relax}
\providecommand{\BIBforeignlanguage}[2]{{%
\expandafter\ifx\csname l@#1\endcsname\relax
\typeout{** WARNING: IEEEtran.bst: No hyphenation pattern has been}%
\typeout{** loaded for the language `#1'. Using the pattern for}%
\typeout{** the default language instead.}%
\else
\language=\csname l@#1\endcsname
\fi
#2}}
\providecommand{\BIBdecl}{\relax}
\BIBdecl

\bibitem{Mahmood2020}
N.~H. {Mahmood}, H.~{Alves}, O.~A. {López}, M.~{Shehab}, D.~P.~M. {Osorio},
  and M.~{Latva-Aho}, ``{Six Key Features of Machine Type Communication in
  6G},'' in \emph{2nd 6G Wireless Summit (6G SUMMIT)}, 2020, pp. 1--5.

\bibitem{David18}
K.~{David} and H.~{Berndt}, ``{6G Vision and Requirements: Is There Any Need
  for Beyond 5G?}'' \emph{{IEEE} Veh. Technol. Mag.}, vol.~13, no.~3, pp.
  72--80, Jul. 2018.

\bibitem{chen2020massive}
\BIBentryALTinterwordspacing
X.~{Chen}, D.~W.~K. {Ng}, W.~{Yu}, E.~G. {Larsson}, N.~{Al-Dhahir}, and
  R.~{Schober}, ``{Massive Access for 5G and Beyond},'' \emph{arXiv e-prints},
  Feb. 2020. [Online]. Available: \url{https://arxiv.org/pdf/2002.03491.pdf}
\BIBentrySTDinterwordspacing

\bibitem{Mumtaz2017}
S.~{Mumtaz}, A.~{Alsohaily}, Z.~{Pang}, A.~{Rayes}, K.~F. {Tsang}, and
  J.~{Rodriguez}, ``{Massive Internet of Things for Industrial Applications:
  Addressing Wireless IIoT Connectivity Challenges and Ecosystem
  Fragmentation},'' \emph{{IEEE} Ind. Electron. Mag.}, vol.~11, no.~1, pp.
  28--33, Mar. 2017.

\bibitem{Mehmood2017}
Y.~{Mehmood}, F.~{Ahmad}, I.~{Yaqoob}, A.~{Adnane}, M.~{Imran}, and
  S.~{Guizani}, ``{Internet-of-Things-Based Smart Cities: Recent Advances and
  Challenges},'' \emph{{IEEE} Commun. Mag.}, vol.~55, no.~9, pp. 16--24, Sep.
  2017.

\bibitem{Nanda2019}
A.~{Nanda}, D.~{Puthal}, J.~J. P.~C. {Rodrigues}, and S.~A. {Kozlov},
  ``{Internet of Autonomous Vehicles Communications Security: Overview, Issues,
  and Directions},'' \emph{{IEEE} Wireless Commun.}, vol.~26, no.~4, pp.
  60--65, Aug. 2019.

\bibitem{Habibzadeh2020}
H.~{Habibzadeh}, K.~{Dinesh}, O.~{Rajabi Shishvan}, A.~{Boggio-Dandry},
  G.~{Sharma}, and T.~{Soyata}, ``{A Survey of Healthcare Internet of Things
  (HIoT): A Clinical Perspective},'' \emph{{IEEE} Internet Things J.}, vol.~7,
  no.~1, pp. 53--71, Jan. 2020.

\bibitem{Ding2017}
Z.~Ding, X.~Lei, G.~K. Karagiannidis, R.~Schober, J.~Yuan, and V.~K. Bhargava,
  ``{A Survey on Non-Orthogonal Multiple Access for 5G Networks: Research
  Challenges and Future Trends},'' \emph{{IEEE} J. Sel. Areas Commun.},
  vol.~35, no.~10, pp. 2181--2195, Oct. 2017.

\bibitem{Chen18}
X.~{Chen}, Z.~{Zhang}, C.~{Zhong}, R.~{Jia}, and D.~W.~K. {Ng}, ``{Fully
  Non-Orthogonal Communication for Massive Access},'' \emph{{IEEE} Trans.
  Commun.}, vol.~66, no.~4, pp. 1717--1731, Apr. 2018.

\bibitem{Shin2017}
W.~{Shin}, M.~{Vaezi}, B.~{Lee}, D.~J. {Love}, J.~{Lee}, and H.~V. {Poor},
  ``{Non-Orthogonal Multiple Access in Multi-Cell Networks: Theory,
  Performance, and Practical Challenges},'' \emph{{IEEE} Commun. Mag.},
  vol.~55, no.~10, pp. 176--183, Oct. 2017.

\bibitem{Wan2018}
D.~{Wan}, M.~{Wen}, F.~{Ji}, H.~{Yu}, and F.~{Chen}, ``{Non-Orthogonal Multiple
  Access for Cooperative Communications: Challenges, Opportunities, and
  Trends},'' \emph{{IEEE} Wireless Commun.}, vol.~25, no.~2, pp. 109--117, Apr.
  2018.

\bibitem{Vaezi2019}
M.~{Vaezi}, R.~{Schober}, Z.~{Ding}, and H.~V. {Poor}, ``{Non-Orthogonal
  Multiple Access: Common Myths and Critical Questions},'' \emph{{IEEE}
  Wireless Commun.}, vol.~26, no.~5, pp. 174--180, Oct. 2019.

\bibitem{polyanskiy2010channel}
Y.~Polyanskiy, H.~V. Poor, and S.~Verd{\'u}, ``{Channel Coding Rate in the
  Finite Blocklength Regime},'' \emph{{IEEE} Trans. Inf. Theory}, vol.~56,
  no.~5, pp. 2307--2359, May 2010.

\bibitem{durisi16}
G.~Durisi, T.~Koch, and P.~Popovski, ``{Toward Massive, Ultrareliable, and
  Low-Latency Wireless Communication With Short Packets},'' \emph{Proc. of the
  IEEE}, vol. 104, no.~9, pp. 1711--1726, Sep. 2016.

\bibitem{Liva2019Survey}
M.~C. Co\c{s}kun, G.~Durisi, T.~Jerkovits, G.~Liva, W.~Ryan, B.~Stein, and
  F.~Steiner, ``{Efficient Error-Correcting Codes in the Short Blocklength
  Regime},'' \emph{Elsevier Phys. Commun.}, vol.~34, pp. 66--79, Jun. 2019.

\bibitem{shental2017low}
O.~Shental, B.~M. Zaidel, and S.~Shamai, ``{Low-Density Code-Domain NOMA:
  Better be Regular},'' in \emph{IEEE Int. Symp. on Inf. Theory (ISIT)}, 2017,
  pp. 2628--2632.

\bibitem{Hoshyar2008}
R.~{Hoshyar}, F.~P. {Wathan}, and R.~{Tafazolli}, ``{Novel Low-Density
  Signature for Synchronous CDMA Systems Over AWGN Channel},'' \emph{{IEEE}
  Trans. Signal Process.}, vol.~56, no.~4, pp. 1616--1626, Apr. 2008.

\bibitem{vdBeek2009}
J.~{van de Beek} and B.~M. {Popovic}, ``{Multiple Access with Low-Density
  Signatures},'' in \emph{IEEE Global Telecommunications Conference
  (GLOBECOMM)}, 2009, pp. 1--6.

\bibitem{Hoshyar10}
R.~{Hoshyar}, R.~{Razavi}, and M.~{Al-Imari}, ``{LDS-OFDM an Efficient Multiple
  Access Technique},'' in \emph{IEEE Vehicular Technology Conference (VTC)},
  2010, pp. 1--5.

\bibitem{Nikopur2013}
H.~{Nikopour} and H.~{Baligh}, ``{Sparse Code Multiple Access},'' in \emph{IEEE
  Annual International Symposium on Personal, Indoor, and Mobile Radio
  Communications (PIMRC)}, 2013, pp. 332--336.

\bibitem{Wang1999}
X.~Wang and H.~V. Poor, ``{Iterative (Turbo) Soft Interference Cancellation and
  Decoding for Coded CDMA},'' \emph{{IEEE} Trans. Commun.}, vol.~47, no.~7, pp.
  1046--1061, Jul. 1999.

\bibitem{Xiao2015}
B.~{Xiao}, K.~{Xiao}, S.~{Zhang}, Z.~{Chen}, B.~{Xia}, and H.~{Liu},
  ``{Iterative Detection and Decoding for SCMA Systems with LDPC Codes},'' in
  \emph{International Conference on Wireless Communications Signal Processing
  (WCSP)}, 2015, pp. 1--5.

\bibitem{Wu2015}
Y.~{Wu}, S.~{Zhang}, and Y.~{Chen}, ``{Iterative Multiuser Receiver in Sparse
  Code Multiple Access Systems},'' in \emph{IEEE International Conference on
  Communications (ICC)}, 2015, pp. 2918--2923.

\bibitem{Meng2018}
X.~{Meng}, Y.~{Wu}, C.~{Wang}, and Y.~{Chen}, ``{Turbo-Like Iterative
  Multi-User Receiver Design for 5G Non-Orthogonal Multiple Access},'' in
  \emph{IEEE Vehicular Technology Conference (VTC-Fall)}, 2018, pp. 1--5.

\bibitem{taherzadeh2014scma}
M.~Taherzadeh, H.~Nikopour, A.~Bayesteh, and H.~Baligh, ``{SCMA Codebook
  Design},'' in \emph{IEEE Vehicular Technology Conference (VTC-Fall)}, Dec.
  2014, pp. 1--5.

\bibitem{Vameghestahbanati2019}
M.~{Vameghestahbanati}, I.~D. {Marsland}, R.~H. {Gohary}, and
  H.~{Yanikomeroglu}, ``{Multidimensional Constellations for Uplink SCMA
  Systems — A Comparative Study},'' \emph{{IEEE Commun. Surveys \&
  Tutorials}}, vol.~21, no.~3, pp. 2169--2194, 2019.

\bibitem{Peng2017}
J.~{Peng}, W.~{Chen}, B.~{Bai}, X.~{Guo}, and C.~{Sun}, ``{Joint Optimization
  of Constellation With Mapping Matrix for SCMA Codebook Design},''
  \emph{{IEEE} Signal Process. Lett.}, vol.~24, no.~3, pp. 264--268, Mar. 2017.

\bibitem{polyanskiy17}
Y.~Polyanskiy, ``{A Perspective on Massive Random-Access},'' in \emph{IEEE
  International Symposium on Information Theory (ISIT)}, Jun. 2017, pp.
  2523--2527.

\bibitem{Abramson1970}
N.~Abramson, ``{The ALOHA System: Another Alternative for Computer
  Communications},'' in \emph{ACM Fall Joint Computer Conference (AFIPS)}, Nov.
  1970, p. 281–285.

\bibitem{Ahlswede1973}
R.~Ahlswede, ``{Multi-way Communication Channels},'' in \emph{International
  Symposium on Information Theory}, Sep. 1971, pp. 23--52.

\bibitem{Rimoldi1996}
B.~{Rimoldi} and R.~{Urbanke}, ``{A Rate-Splitting Approach to the Gaussian
  Multiple-Access Channel},'' \emph{{IEEE} Trans. Inf. Theory}, vol.~42, no.~2,
  pp. 364--375, Mar. 1996.

\bibitem{Zhu2017}
J.~{Zhu} and M.~{Gastpar}, ``{Gaussian Multiple Access via
  Compute-and-Forward},'' \emph{{IEEE} Trans. Inf. Theory}, vol.~63, no.~5, pp.
  2678--2695, May 2017.

\bibitem{Mathys1990}
P.~Mathys, ``{A Class of Codes for a T Active Users out of N Multiple-Access
  Communication System},'' \emph{{IEEE} Trans. Inf. Theory}, vol.~36, no.~6,
  pp. 1206--1219, Nov. 1990.

\bibitem{bar-david93}
I.~Bar-David, E.~Plotnik, and R.~Rom, ``{Forward Collision Resolution - a
  Technique for Random Multiple-Access to the Adder Channel},'' \emph{{IEEE}
  Trans. Inf. Theory}, vol.~39, no.~5, pp. 1671--1675, Sep. 1993.

\bibitem{Casini2007}
E.~{Casini}, R.~{De Gaudenzi}, and O.~{Del Rio Herrero}, ``{Contention
  Resolution Diversity Slotted ALOHA (CRDSA): An Enhanced Random Access Scheme
  for Satellite Access Packet Networks},'' \emph{{IEEE} Trans. Wireless
  Commun.}, vol.~6, no.~4, pp. 1408--1419, Apr. 2007.

\bibitem{paolini2015bcoded}
E.~Paolini, C.~Stefanovic, G.~Liva, and P.~Popovski, ``{Coded Random Access:
  Applying Codes on Graphs to Design Random Access Protocols},'' \emph{{IEEE}
  Commun. Mag.}, vol.~53, no.~6, pp. 144--150, Jun. 2015.

\bibitem{chen14}
X.~Chen and D.~Guo, ``{Many-Access Channels: The Gaussian Case with Random user
  Activities},'' in \emph{IEEE International Symposium on Information Theory
  (ISIT)}, Jun. 2014, pp. 3127--3131.

\bibitem{Chen2017}
X.~{Chen}, T.~{Chen}, and D.~{Guo}, ``{Capacity of Gaussian Many-Access
  Channels},'' \emph{{IEEE} Trans. Inf. Theory}, vol.~63, no.~6, pp.
  3516--3539, Feb. 2017.

\bibitem{Ravi2020}
J.~{Ravi} and T.~{Koch}, ``{Capacity per Unit-Energy of Gaussian Random
  Many-Access Channels},'' in \emph{IEEE International Symposium on Information
  Theory (ISIT)}, 2020, pp. 3025--3030.

\bibitem{shamai97}
S.~Shamai, ``{A Broadcast Strategy for the Gaussian slowly fading Channel},''
  in \emph{IEEE International Symposium on Information Theory (ISIT)}, Jun.
  1997, pp. 150--.

\bibitem{ordentlich2017low}
O.~Ordentlich and Y.~Polyanskiy, ``{Low Complexity Schemes for the Random
  Access Gaussian Channel},'' in \emph{IEEE International Symposium on
  Information Theory (ISIT)}, Jun. 2017, pp. 2528--2532.

\bibitem{yoshida2006analysis}
M.~Yoshida and T.~Tanaka, ``{Analysis of Sparsely-Spread CDMA via Statistical
  Mechanics},'' in \emph{2006 IEEE International Symposium on Information
  Theory (ISIT)}, Seattle, Washington, Dec. 2006, pp. 2378--2382.

\bibitem{verdu1999spectral}
S.~Verd{\'u} and S.~Shamai, ``{Spectral Efficiency of CDMA with Random
  Spreading},'' \emph{{IEEE} Trans. Inf. Theory}, vol.~45, no.~2, pp. 622--640,
  Mar. 1999.

\bibitem{Zaidel2018}
B.~M. {Zaidel}, O.~{Shental}, and S.~S. {Shitz}, ``{Sparse NOMA: A Closed-Form
  Characterization},'' in \emph{IEEE International Symposium on Information
  Theory (ISIT)}, 2018, pp. 1106--1110.

\bibitem{Ferrante18}
M.~T.~P. {Le}, G.~C. {Ferrante}, T.~Q.~S. {Quek}, and M.~{Di Benedetto},
  ``{Fundamental Limits of Low-Density Spreading NOMA With Fading},''
  \emph{{IEEE} Trans. Wireless Commun.}, vol.~17, no.~7, pp. 4648--4659, Jul.
  2018.

\bibitem{GallagerLDPC}
R.~{Gallager}, ``{Low-Density Parity-Check Codes},'' \emph{IRE Trans. on Inf.
  Theory}, vol.~8, no.~1, pp. 21--28, Jan. 1962.

\bibitem{XIAOHU2005}
{Xiao-Yu Hu}, E.~{Eleftheriou}, and D.~M. {Arnold}, ``{Regular and Irregular
  Progressive Edge-Growth Tanner Graphs},'' \emph{{IEEE} Trans. Inf. Theory},
  vol.~51, no.~1, pp. 386--398, Jan. 2005.

\bibitem{Qi2017}
T.~{Qi}, W.~{Feng}, Y.~{Chen}, and Y.~{Wang}, ``{When NOMA Meets Sparse Signal
  Processing: Asymptotic Performance Analysis and Optimal Sequence Design},''
  \emph{{IEEE} Access}, vol.~5, pp. 18\,516--18\,525, Jul. 2017.

\bibitem{Qi2017b}
T.~{Qi}, W.~{Feng}, and Y.~{Wang}, ``{Optimal Sequences for Non-Orthogonal
  Multiple Access: A Sparsity Maximization Perspective},'' \emph{{IEEE} Commun.
  Lett.}, vol.~21, no.~3, pp. 636--639, Mar. 2017.

\bibitem{liva11}
G.~Liva, ``{Graph-Based Analysis and Optimization of Contention Resolution
  Diversity Slotted ALOHA},'' \emph{{IEEE} Trans. Commun.}, vol.~59, no.~2, pp.
  477--487, Feb. 2011.

\bibitem{Boyd2019}
C.~{Boyd}, R.~{Kotaba}, O.~{Tirkkonen}, and P.~{Popovski}, ``{Non-Orthogonal
  Contention-Based Access for URLLC Devices with Frequency Diversity},'' in
  \emph{IEEE Int. Workshop on Signal Processing Advances in Wireless
  Communications (SPAWC)}, 2019, pp. 1--5.

\bibitem{Ustinova2019}
D.~{Ustinova}, A.~{Glebov}, P.~{Rybin}, and A.~{Frolov}, ``{Efficient
  Concatenated Same Codebook Construction for the Random Access Gaussian
  MAC},'' in \emph{IEEE Vehicular Technology Conference (VTC2019-Fall)}, 2019,
  pp. 1--5.

\bibitem{Fengler2019}
A.~{Fengler}, P.~{Jung}, and G.~{Caire}, ``{SPARCs and AMP for Unsourced Random
  Access},'' in \emph{IEEE International Symposium on Information Theory
  (ISIT)}, 2019, pp. 2843--2847.

\bibitem{Fengler2020}
------, ``{Unsourced Multiuser Sparse Regression Codes achieve the Symmetric
  MAC Capacity},'' in \emph{IEEE International Symposium on Information Theory
  (ISIT)}, 2020, pp. 3001--3006.

\bibitem{Amalladinne2020}
V.~K. {Amalladinne}, J.~F. {Chamberland}, and K.~R. {Narayanan}, ``{A Coded
  Compressed Sensing Scheme for Unsourced Multiple Access},'' \emph{{IEEE}
  Trans. Inf. Theory}, vol.~66, no.~10, pp. 6509--6533, Jul. 2020.

\bibitem{Dommel19}
J.~{Dommel}, Z.~{Utkovski}, L.~{Thiele}, and S.~{Stańczak}, ``{Sparse
  Code-Domain Non-Orthogonal Random Access with Peeling Decoder},'' in
  \emph{IEEE Asilomar Conference on Signals, Systems, and Computers}, 2019, pp.
  984--988.

\bibitem{Luby2001}
M.~G. {Luby}, M.~{Mitzenmacher}, M.~A. {Shokrollahi}, and D.~A. {Spielman},
  ``{Efficient Erasure Correcting Codes},'' \emph{{IEEE} Trans. Inf. Theory},
  vol.~47, no.~2, pp. 569--584, Feb. 2001.

\bibitem{Liu2019}
Y.~{Liu}, P.~M. {Olmos}, and T.~{Koch}, ``{A Probabilistic Peeling Decoder to
  Efficiently Analyze Generalized LDPC Codes Over the BEC},'' \emph{{IEEE}
  Trans. Inf. Theory}, vol.~65, no.~8, pp. 4831--4853, Aug. 2019.

\bibitem{zeng2016peeling}
W.~Zeng and H.~Wang, ``{Peeling Decoding of LDPC Codes with Applications in
  Compressed Sensing},'' \emph{Mathematical Problems in Engineering}, Apr.
  2016.

\bibitem{Li19}
X.~{Li}, D.~{Yin}, S.~{Pawar}, R.~{Pedarsani}, and K.~{Ramchandran},
  ``{Sub-Linear Time Support Recovery for Compressed Sensing Using Sparse-Graph
  Codes},'' \emph{{IEEE} Trans. Inf. Theory}, vol.~65, no.~10, pp. 6580--6619,
  Oct. 2019.

\bibitem{macneish1922euler}
H.~F. MacNeish, ``{Euler Squares},'' \emph{Princeton University Annals of
  Mathematics}, vol.~23, no.~3, pp. 221--227, Mar. 1922.

\bibitem{Naidu2016}
R.~R. Naidu, P.~Jampana, and C.~S. Sastry, ``{Deterministic Compressed Sensing
  Matrices: Construction via Euler Squares and Applications},'' \emph{{IEEE}
  Trans. Signal Process.}, vol.~64, no.~14, pp. 3566--3575, Jul. 2016.

\bibitem{wallis2016introduction}
W.~D. Wallis and J.~C. George, \emph{{Introduction to Combinatorics}}.\hskip
  1em plus 0.5em minus 0.4em\relax CRC press, 2016.

\bibitem{Boyd18}
C.~{Boyd}, R.~{Vehkalahti}, and O.~{Tirkkonen}, ``{Grant-Free Access in URLLC
  with Combinatorial Codes and Interference Cancellation},'' in \emph{IEEE
  Globecom Workshops (GC Wkshps)}, Dec. 2018, pp. 1--5.

\bibitem{Diao16}
Q.~{Diao}, J.~{Li}, S.~{Lin}, and I.~F. {Blake}, ``{New Classes of Partial
  Geometries and their Associated LDPC Codes},'' \emph{{IEEE} Trans. Inf.
  Theory}, vol.~62, no.~6, pp. 2947--2965, Jun. 2016.

\bibitem{Fang2015}
Y.~{Fang}, G.~{Bi}, Y.~L. {Guan}, and F.~C.~M. {Lau}, ``{A Survey on Protograph
  LDPC Codes and Their Applications},'' \emph{IEEE Commun. Surveys \&
  Tutorials}, vol.~17, no.~4, pp. 1989--2016, May 2015.

\bibitem{Divsalar2009}
D.~{Divsalar}, S.~{Dolinar}, C.~R. {Jones}, and K.~{Andrews}, ``{Capacity -
  Approaching Protograph Codes},'' \emph{{IEEE} J. Sel. Areas Commun.},
  vol.~27, no.~6, pp. 876--888, Jul. 2009.

\bibitem{Abbasfar2007}
A.~{Abbasfar}, D.~{Divsalar}, and K.~{Yao}, ``{Accumulate-Repeat-Accumulate
  Codes},'' \emph{{IEEE} Trans. Commun.}, vol.~55, no.~4, pp. 692--702, Apr.
  2007.

\bibitem{Johnson2004}
S.~J. {Johnson} and S.~R. {Weller}, ``{Codes for Iterative Decoding from
  Partial Geometries},'' \emph{{IEEE} Trans. Commun.}, vol.~52, no.~2, pp.
  236--243, Feb. 2004.

\bibitem{Pradhan2019}
A.~K. {Pradhan}, V.~K. {Amalladinne}, K.~R. {Narayanan}, and J.~{Chamberland},
  ``{Polar Coding and Random Spreading for Unsourced Multiple Access},'' in
  \emph{IEEE International Conference on Communications (ICC)}, 2020, pp. 1--6.

\bibitem{Marshakov2019}
E.~{Marshakov}, G.~{Balitskiy}, K.~{Andreev}, and A.~{Frolov}, ``{A Polar Code
  Based Unsourced Random Access for the Gaussian MAC},'' in \emph{IEEE
  Vehicular Technology Conference (VTC2019-Fall)}, Sep. 2019, pp. 1--5.

\bibitem{Ebada2018}
M.~Ebada, A.~Elkelesh, S.~Cammerer, and S.~ten Brink, ``{Scattered EXIT Charts
  for Finite Length LDPC Code Design},'' in \emph{IEEE International Conference
  on Communications (ICC)}, May 2018, pp. 1--7.

\end{thebibliography}
\bibliographystyle{IEEEtran}

\end{document}